%% file: main.tex
\documentclass[letterpaper,twocolumn,10pt]{article}
\usepackage{usenix-2020-09}
\usepackage{minted}
\usepackage{xurl}
\input{packages} 

\begin{document}
\date{}

\title{\TITLE}

\author{
    {\rm Osayamen Jonathan Aimuyo}\\
  Stanford University\\
  {\rm osayamen@stanford.edu}\\
  \and
  {\rm Swapnil Gandhi}\\
  Stanford University\\
  {\rm gandhis@stanford.edu}
  \and
  {\rm Christos Kozyrakis}\\
  NVIDIA \& Stanford University\\
  {\rm kozyraki@stanford.edu}
} 

\maketitle  

\ifthenelse{\equal{\PAGENUMBERS}{no}}{%
  \thispagestyle{empty}
}

\makeatletter
\def\blfootnote{\xdef\@thefnmark{}\@footnotetext}
\makeatother

\newcommand{\todo}[1]{\textcolor{red}{TODO: #1}}

\input{sections/0_abstract}

\input{sections/1_introduction}
\input{sections/2_background}
\input{sections/3_design}
\input{sections/4_implementation}
\input{sections/5_evaluation}
\input{sections/6_discussion}
\input{sections/7_conclusion}

\phantomsection
\label{EndOfPaper}



\bibliographystyle{plain}
\bibliography{references}

\clearpage      
\nobalance      
\input{sections/appendix}

\end{document}

%% file: packages.tex
\usepackage{xspace}

\newcommand{\AUTHORS}{Authors}

\newcommand{\NAME}{Purlin\xspace}

\newcommand{\ATOMTT}{\texttt{Atom}\xspace}

\newcommand{\TITLE}{\NAME: Separating Orchestration from the Datapath of Collectives}
\newcommand{\KEYWORDS}{}
\newcommand{\CONFERENCE}{}
\newcommand{\PAGENUMBERS}{yes}  
\newcommand{\COMMENTS}{yes}      

\newcommand{\para}[1]{\medskip\noindent\textbf{#1}}

\usepackage{balance, ifthen, multirow, times}

\usepackage{inconsolata}

\usepackage{titling}            
\usepackage[small,compact]{titlesec}         

\usepackage[font=small, labelfont=bf]{caption}

\usepackage{subcaption}

\usepackage{fancyhdr}

\ifthenelse{\equal{\PAGENUMBERS}{yes}}{%
  \pagestyle{plain}
}{%
  \pagestyle{empty}
}

\usepackage[numbers,sort&compress]{natbib}

\usepackage{enumitem}
\setlist{itemsep=0pt,parsep=0pt,topsep=0pt} 

\usepackage{booktabs}
\usepackage{colortbl}
\usepackage{float}                           

\usepackage{pgfplots}
\usepackage{array}
\usepackage{graphicx}
\pgfplotsset{compat=1.18}

\newcommand*{\MinNumber}{0.1}%
\newcommand*{\MidNumber}{0.85} %
\newcommand*{\MaxNumber}{1.0}%

\newcommand{\ApplyGradient}[1]{%
        \ifdim #1 pt > \MidNumber pt
            \pgfmathsetmacro{\PercentColor}{max(min(100.0*(#1 - \MidNumber)/(\MaxNumber-\MidNumber),100.0),0.00)} %
            \hspace{-0.33em}\colorbox{green!\PercentColor!yellow}{#1}
        \else
            \pgfmathsetmacro{\PercentColor}{max(min(100.0*(\MidNumber - #1)/(\MidNumber-\MinNumber),100.0),0.00)} %
            \ifdim #1 pt < 0.65 pt 
                \hspace{-0.33em}\colorbox{red!\PercentColor!yellow}{\color{white}{#1}}
            \else
                \hspace{-0.33em}\colorbox{red!\PercentColor!yellow}{#1}
            \fi
        \fi
}

\usepackage{makecell}
\usepackage[dvipsnames]{xcolor}
\usepackage{pifont}
\newcommand{\greencheck}{\color{ForestGreen}{\pmb{\ding{51}}}}
\newcommand{\redcross}{\color{red}{\pmb{\ding{55}}}}
\newcommand{\partialcheck}{{\color{orange}\pmb{\ding{51}}}}

\definecolor{placeholderbg}{rgb}{0.85,0.85,0.85}

\usepackage{mathtools}
\usepackage{amssymb}
\usepackage{pifont}

\usepackage{url}
\hypersetup{%
    pdfauthor = {\AUTHORS},
    pdftitle = {\TITLE},
    pdfsubject = {\CONFERENCE},
    pdfkeywords = {\KEYWORDS},
    bookmarksopen = {true}
}
\usepackage{cleveref}               
\crefname{section}{\S}{\SS}
\crefname{figure}{Figure}{Figures}
\crefname{table}{Table}{Tables}
\crefname{equation}{Eq.}{Eqs.}
\Crefname{equation}{Equation}{Equations}
\crefformat{section}{\S#2#1#3}      
\crefformat{subsection}{\S#2#1#3}
\crefformat{subsubsection}{\S#2#1#3}

\usepackage{listings}
\newcommand\code[1]{\lstinline$#1$}

\usepackage{algorithm}
\usepackage{algpseudocode}
\usepackage{multirow}
\usepackage{pifont}
\usepackage[normalem]{ulem}

\definecolor{string-color}{rgb}{0.3333, 0.5254, 0.345}
\definecolor{commentcolor}{HTML}{5d6063}
\lstdefinestyle{pseudocode}{
    language=Python,
    mathescape=true,
    keywords={FindWindowSize, GenerateSchedule, OrderOperators, SparseCheckpointSchedule, profiler},
    keywordstyle=\color{blue}\bfseries,
    morekeywords={[2]},
    keywordstyle={[2]\bfseries},
    morekeywords={[3]if,def,return,else,while,for,in,append,ceil,break},
    keywordstyle={[3]\bfseries},
    basicstyle=\small\ttfamily,
    identifierstyle=\color{black},
    sensitive=false,
    commentstyle=\color{commentcolor},
    columns=fullflexible,
    keepspaces=false,
    breaklines=true,
    showstringspaces=false,
    numbers=left,
    numberstyle=\tiny,
    numbersep=4pt,
    xleftmargin=5pt,
    lineskip={-1pt},
    postbreak=\mbox{\textcolor{black}{$\hookrightarrow$}\space},
    stringstyle=\color{string-color},
}



\algnewcommand{\LineComment}[1]{\State \(\Comment\) #1}

\usepackage{tikz}
\usetikzlibrary{arrows.meta,positioning}

\DeclareRobustCommand\step[1]{\tikz[baseline=(c.base)]{
    \node[circle,draw,fill=gray!25,inner sep=1pt] (c) {#1};}}

\usepackage{comment} 
\usepackage{soul}
\soulregister\cite7
\soulregister\ref7
\soulregister\pageref7

\usepackage[all]{nowidow}

\ifthenelse{\equal{\COMMENTS}{yes}}{%
    \newcommand\jonathan[1]{\textcolor{magenta}{[Jonathan:] #1}}
    \newcommand\swapnil[1]{\textcolor{blue}{[Swapnil:] #1}}
    \newcommand\christos[1]{\textcolor{red}{[Christos:] #1}}
}{
    \newcommand\jonathan[1]{\unskip}
    \newcommand\swapnil[1]{\unskip}
    \newcommand\christos[1]{\unskip}
}

%% file: sections/0_abstract.tex
\begin{abstract}
Distributed inference depends on GPU collective communication that must keep pace with evolving hardware and specialized workloads.
However, existing works often couple the orchestration (where and when data moves) and datapath (how data moves) of collective communication.
This coupling makes it costly to adopt new hardware mechanisms and customize communication for applications.

We present \NAME\footnote{Code is available at: \url{https://github.com/purlin-project/purlin}}\textsuperscript{,}\footnote{Correspondence to osayamen@stanford.edu}, a scale-up communication framework that separates these concerns.
At the top of \NAME, we specify collectives as a \emph{transformation} from an input to an output \emph{layout}.
In the middle, we introduce a shared protocol, \textbf{S}tage, \textbf{N}otify, \textbf{A}nd \textbf{C}onsume (SNAC), which derives orchestration from these specifications.
Below SNAC sits a hardware-aware datapath we call \texttt{Atom}, which implements two key data movement primitives for collectives: \texttt{copy} and \texttt{reduce}.
This separation lets us customize collectives and adopt new hardware mechanisms while reusing orchestration via SNAC.

We evaluate \NAME on A100, H200, and B200 GPUs. Across seven collectives, \NAME achieves latency speedups of up to 5.14$\times$ and bandwidth improvements of up to 4.50$\times$ over baselines.
Integrated into SGLang, \NAME improves offline LLM serving throughput and interactivity by 1.13$\times$ on average and up to 1.37$\times$ over baselines.
For online LLM inference, \NAME improves interactivity by 1.26$\times$ on average and up to 2.85$\times$, with the largest gain occurring under overload.
For diffusion image generation, \NAME reduces end-to-end latency by up to 1.13$\times$.
\end{abstract}

%% file: sections/1_introduction.tex
\section{Introduction}\label{sec:introduction}
LLMs with hundreds of billions to trillions of parameters require distributed execution to sidestep memory constraints and achieve low latency or high throughput~\cite{inkling2026misc, kimik26misc}.
In this paper, we focus on \emph{scale-up} domains, where GPUs are completely connected over a high-bandwidth interconnect such as NVLink.
Within such a domain, inference frameworks such as vLLM~\cite{vllm}, SGLang~\cite{sglang}, and TensorRT-LLM~\cite{tensorrt} distribute computation through tensor~\cite{megatronlm}, expert~\cite{deepep2025}, and other forms of parallelism~\cite{ulysses, longserve, splitwise, distserve, megascaleinfer}.
To exchange intermediate results, these parallelism strategies rely on \emph{collective communication}
provided by frameworks like NCCL~\cite{11244782, nvidia2025nccl} and other systems~\cite{mscclpp, ncclx, parallelkittens}.

\begin{table}[!t]
  \centering
  \footnotesize
  \setlength{\tabcolsep}{5pt}
  \begin{tabular}{lcccc}
    \toprule
    System & Compat. & Evolv. & Deriv. & Prog. \\
    \midrule
    NCCL~\cite{nvidia2025nccl, 11244782} & \greencheck & \redcross & \redcross & \greencheck \\
    NVSHMEM~\cite{nvidia2026nvshmem} & \greencheck & \redcross & \redcross & \partialcheck \\
    MSCCL++~\cite{mscclpp} & \greencheck & \partialcheck & \redcross & \greencheck \\
    NCCLX~\cite{ncclx} & \greencheck & \redcross & \redcross & \redcross \\
    ParallelKittens~\cite{parallelkittens} & \redcross & \redcross & \redcross & \greencheck \\
    \midrule
    \NAME & \greencheck & \greencheck & \greencheck & \greencheck \\
    \bottomrule
  \end{tabular}
  \caption{Properties of scale-up communication systems. \emph{Compatibility}: supports Ampere, Hopper, and Blackwell, as well as older GPUs that rely on vectorized loads and stores. \emph{Evolvability}: can evolve the datapath of collectives without rewriting orchestration and vice versa. \emph{Derivability}: derives orchestration for collectives from their semantics. \emph{Programmability}: exposes interfaces for customizing the datapath or orchestration of collectives.}
  \label{tab:properties}
\end{table}

However, attaining high performance and \emph{adaptability} within a collective communication framework is challenging because both hardware and application requirements constantly change.
On the one hand, new GPU generations continue to increase interconnect bandwidth and rapidly introduce hardware mechanisms that accelerate data movement~\cite{nvidia2020ampere,hopper_whitepaper}.
On the other, applications require different communication regimes, from throughput-oriented prefill to latency-sensitive decode, as well as device-resident building blocks that can fuse with computation~\cite{318585, NEURIPS2025_918d938b}.
Serving these demands requires an implementation that lets collectives and hardware mechanisms \emph{evolve independently} without sacrificing performance.

\para{Coupled, inflexible collectives.}
A collective's \emph{semantics} specify input and output contracts for all participating ranks as formalized by MPI~\cite{mpi10journal}.
Its \emph{orchestration} determines where and when data moves, specifically the communication schedule and inter-rank synchronization,
while its \emph{datapath} implements the copies or reductions that make up the collective using available hardware mechanisms.
Existing systems expose reusable GPU datapath primitives~\cite{11244782, mscclpp,parallelkittens, ncclx}, but they often implement collectives by coupling orchestration and the corresponding datapath.
This coupling makes it difficult to evolve these implementations \emph{selectively}, as is often needed when tuning for an application-specific use case or adopting a new hardware mechanism or a communication pattern not covered by provided primitives.
We argue for decoupling orchestration from the datapath of collectives as a path towards making GPU communication flexible and evolvable without sacrificing performance or portability.
We elaborate on the problems motivating this decoupling below.

\para{Problem 1: Evolvability and compatibility are challenging to retain together.}
When collective communication orchestration is coupled with its datapath, adopting a new transfer mechanism requires modifications beyond the datapath alone.
This coupling increases the effort to adopt new mechanisms while retaining support for older, still deployed GPUs.
As a result, within existing communication frameworks, support for new hardware mechanisms lags behind, highlighting insufficient \emph{evolvability}.
For instance, an established communication library~\cite{nvidia2026nvshmem} incorporated the asynchronous Tensor Memory Accelerator (TMA) into its datapath in 2026, four years after Hopper introduced the mechanism~\cite{nvshmem_rn_370}.
Newer works~\cite{parallelkittens} point out this gap but end up specializing in a small set of generations like Hopper or Blackwell without supporting still deployed but older hardware like Ampere.
This illustrates the challenge of maintaining \emph{compatibility} with older deployed GPUs while attaining evolvability.

\para{Problem 2: New collectives still require explicit orchestration.}
Applications need collectives and variants that existing libraries do not provide or do not optimize sufficiently.
Application developers often fill this gap with custom implementations, including AllReduce in vLLM~\cite{hanzhi2024customallreduce}, SGLang~\cite{sgljitar}, and TensorRT-LLM~\cite{korzh2024multishot}.
To build such collectives, developers leverage datapath and synchronization primitives exposed by device-side APIs in NVSHMEM, NCCL, or MSCCL++~\cite{nvidia2026nvshmem,jeaugey2025nccl228,mscclpp}.
However, implementing a new collective in these frameworks still requires translating its semantics into orchestration.
What is missing is \emph{derivability}: the ability to specify a collective's semantics and have the framework derive efficient orchestration.

\para{Problem 3: Customization can require revisiting orchestration.}
Applications need to customize how communication executes, including controlling reduction order~\cite{mai_thinking_1}, tuning transfers for particular hardware, and fusing communication with computation~\cite{NEURIPS2025_918d938b,318585}.
Existing systems support such customization through device-side primitives and programmable interfaces~\cite{nvidia2026nvshmem,jeaugey2025nccl228,mscclpp,parallelkittens}.
However, when a collective's orchestration is coupled with its data movement, modifying the datapath can also require revisiting orchestration.
Our goal is \emph{programmability} with reusable orchestration: we aim to specialize communication components within a collective without reimplementing orchestration logic from scratch.

\para{Where we are today.} \Cref{tab:properties} summarizes our observations of existing systems. NCCL~\cite{nvidia2025nccl}, NCCLX~\cite{ncclx}, and MSCCL++~\cite{mscclpp} are portable but provide coupled collectives that are inconvenient to evolve. NVSHMEM~\cite{nvidia2026nvshmem}, MSCCL++, and, recently, NCCL's device API~\cite{jeaugey2025nccl228} additionally offer programmable datapath primitives, but neither provides a mechanism for \emph{deriving} high-performance collectives.
ParallelKittens~\cite{parallelkittens} specializes for more recent GPU generations by tightly coupling to their datapaths, but in doing so sacrifices compatibility with prior generations, which are still widely deployed today.
Together, we observe a spectrum with \textit{portable but challenging to evolve} systems at one end and \textit{specialized but generation bound} implementations at the other.

\para{Where we should be.}
Can a communication framework be compatible across hardware generations, convenient to evolve and adaptable to evolving workloads and communication patterns, without sacrificing performance?
We argue it can.
Our \textbf{key observation} is that the orchestration and datapath of collectives evolve along \emph{different axes}: orchestration evolves primarily with application characteristics, while the datapath evolves with hardware generations.
Therefore, we propose \emph{separating} these two concerns into independent layers that interoperate within a unified communication framework.
This separation of concerns, inspired by prior work~\cite{clark1988design, engler1995exokernel, akhshabi2011hourglass, lattner2004llvm, ragankelley2013halide}, lets us evolve either orchestration or the datapath independently as we keep pace with application and hardware evolution.

\begin{figure}[!t]
  \centering
  \includegraphics[width=0.4\textwidth, keepaspectratio]{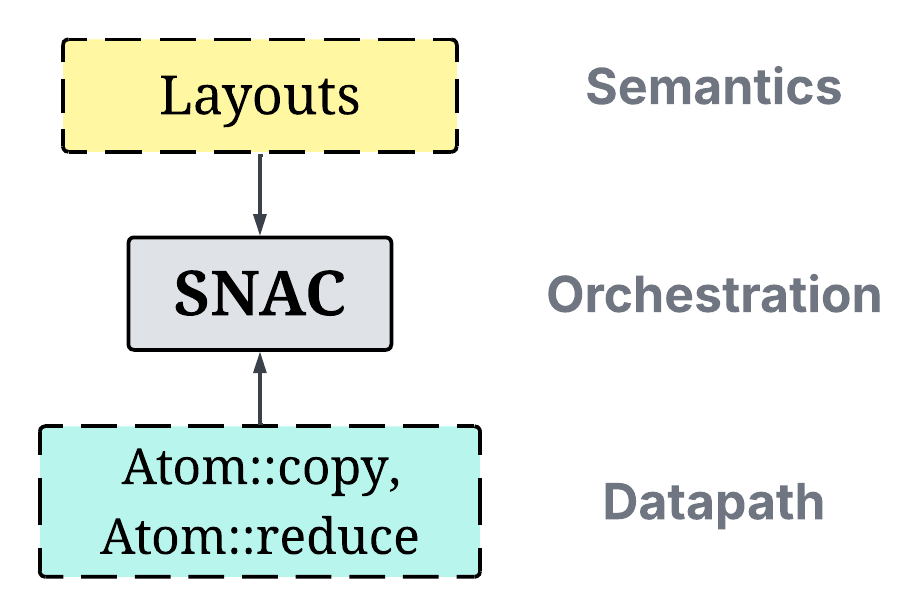}
  \caption{\NAME's decoupled stack. We use layouts to describe \emph{what} collective to execute. SNAC orchestrates \emph{where and when} data moves in this collective, while the \texttt{Atom} deals with \emph{how} data moves on a specific GPU generation.}
  \label{fig:tiers}
\end{figure}

\para{Our design.}
We present \NAME\footnote{A purlin supports a roof covering (above) on a frame that sits below. Similarly, \NAME provides an orchestration layer between collective communication semantics (above) and the hardware datapath (below).}, a communication framework for the scale-up domain whose key innovation is \textbf{decoupling} orchestration from the datapath of collectives (\cref{fig:tiers}). At the top, we describe collectives as transformations over a concise family of layouts (Listing~\ref{lst:collective-namings} and \cref{subsec:layouts}).
Our middle layer derives orchestration through the \textbf{S}tage, \textbf{N}otify, \textbf{A}nd \textbf{C}onsume (SNAC) protocol (\cref{subsec:snac}).
SNAC controls the communication schedule and dispatches data transfers to the bottom layer, the \texttt{Atom}, which exposes hardware-aware primitives for copies and reduction (\cref{subsec:atom}).

\para{Why this separation is hard.}
\step{1} Collectives differ vastly in their input and output semantics.
Variable-length collectives also impose non-uniform partitioning across ranks.
We capture these semantics with a concise set of \emph{layouts}, enabling SNAC to orchestrate collectives through \emph{shared} layout rules (\cref{subsec:layouts}).
\step{2} Within each collective, latency-bound~\cite{shen2026micro} and throughput-bound transfers demand different execution strategies.
We serve both regimes through specializations of the same SNAC protocol (\cref{subsec:snac}, \cref{fig:snac-execution-paths}).
\step{3} Supporting multiple layouts and datapaths could introduce runtime overhead if done \emph{dynamically}.
Thus, we use template metaprogramming~\cite{vandevoorde2017templates} in SNAC to \emph{statically} derive orchestration from layouts and to select Atom implementations at compile time (\cref{subsec:atom}). \step{4} Separating orchestration from the datapath of collectives risks losing the \emph{codesign} between algorithm and hardware that yields desirable performance.
We therefore configure SNAC via \emph{codesign policies}, compile-time parameters that users can tune per GPU generation or communication use case (\cref{subsec:implementation-policies}).

By meeting these challenges, our stack maintains the following properties.

\para{Compatibility.}
The \ATOMTT layer ensures hardware compatibility by construction as it captures the details of the executing hardware.
Concretely, \NAME provides a baseline and broadly compatible \ATOMTT using only vectorized load and store instructions, in addition to an Ampere-specialized \ATOMTT incorporating asynchronous data movement instructions and even more specialized Hopper and Blackwell Atoms for improved performance~(\S\ref{subsec:atom}, \S\ref{sec:implementation}).

\para{Evolvability.}
In \NAME, adopting a new datapath means implementing a new \ATOMTT while extending orchestration confines changes to SNAC only.
We evaluate the performance achieved by this separation across three hardware generations through collective microbenchmarks (\cref{subsec:microbenchmarks}) and end-to-end inference evaluations (\cref{subsec:application-performance}).

\para{Derivability.}
\NAME derives orchestration from a high-level description.
Through this technique, we express seven non-rooted collectives (\cref{subsec:layouts}), including variable-length variants, and show how to express four more rooted collectives in \cref{app:rooted-collectives}.
We evaluate our collectives in \cref{subsec:microbenchmarks} and \cref{subsec:varlen-collectives}.

\para{Programmability.}
SNAC and the \ATOMTT are composable building blocks enabling \NAME collectives to be flexible by construction.
Specifically, a user can easily extend the \ATOMTT or develop a codesign policy for SNAC to meet their needs.
We demonstrate the performance impact of this expressiveness via ablations in \S\ref{subsec:ablation}.

\para{Results.}
We run microbenchmarks (\cref{subsec:microbenchmarks}) and end-to-end inference (\cref{subsec:application-performance}) evaluations on A100s, H200s, and B200s.
Across seven collectives, we achieve latency speedups of up to 5.14$\times$ and bandwidth improvements of up to 4.50$\times$ over existing baselines.
In SGLang, we improve offline LLM serving interactivity by 1.13$\times$ on average and up to 1.37$\times$ across three GPU generations.
We improve online LLM interactivity by 1.26$\times$ on average and up to 2.85$\times$.
We also speed up diffusion image generation by up to 1.13$\times$.

%% file: sections/2_background.tex
\section{Background and Motivation}\label{sec:background}
In this section, we review collective communication semantics (\cref{subsec:collective-semantics}) and categorize what requirements evolving applications (\cref{subsec:evolving-applications}) and hardware (\cref{subsec:evolving-hardware}) impose on a communication framework.
\begin{table}[!ht]
  \centering
  \footnotesize
  \setlength{\tabcolsep}{3pt}
  \begin{tabular}{@{}>{\raggedright\arraybackslash}p{0.20\columnwidth}>{\raggedright\arraybackslash}p{0.39\columnwidth}>{\raggedright\arraybackslash}p{0.35\columnwidth}@{}}
    \toprule
    Parallelism & Collectives & Function \\
    \midrule
    Tensor~\cite{megatronlm} & AllReduce & Aggregate activations \\
    \midrule
    Expert~\cite{deepep2025} & AllGather(V), ReduceScatter(V), AllToAll(V) & Dispatch tokens and combine expert outputs \\
    \midrule
    Data parallel attention~\cite{sglang} & AllGather(V), ReduceScatter(V) & Exchange activations between attention and FFN or MoE layers \\
    \midrule
    Ulysses sequence~\cite{ulysses} & AllToAll & Redistribute a partitioned sequence \\
    \bottomrule
  \end{tabular}
  \caption{Collective communication in distributed inference.}
  \label{tab:background-parallelism}
\end{table}
\subsection{Collective communication semantics}\label{subsec:collective-semantics}
Collective communication arises in inference from model parallelism strategies.
Specifically, \cref{tab:background-parallelism} shows these parallelism strategies and the collective communication they invoke.
The semantics for these collectives, as standardized by MPI~\cite{mpi10journal} three decades ago, specify input preconditions and output postconditions across all participating ranks.
These contracts are \emph{stable} and independent of the datapath in which an implementation realizes collectives.
That is, communication frameworks, for instance, may implement AllReduce through various algorithms; however, all such implementations must adhere to AllReduce’s (1) precondition of each rank providing an input buffer holding a contribution and (2) postcondition of an output buffer holding the accumulation of all such input contributions.
We therefore consider these semantics, specifically the buffer contracts, a useful basis for a \emph{high-level} and general description of collectives independent of implementation details.

\subsection{Evolving applications}\label{subsec:evolving-applications}
\input{figures/purlin_breakdown_fig}
\para{Demand for regime-aware orchestration.}
Applications increasingly have heterogeneous execution phases that demand different techniques to sustain peak performance.
A relevant example is LLM inference, where the prefill phase processes the entire sequence in one pass, producing communication payloads of tens to hundreds of MB (large sizes).
In contrast, the decode phase generates one token at a time, resulting in payloads sized tens to hundreds of KB (small sizes)~\cite{shen2026micro}.
A communication framework catering to LLM inference must therefore \emph{specialize its orchestration} for both regimes since
throughput-bound techniques for large input sizes impose costs that otherwise dominate latency at smaller input sizes.
The converse also holds: latency-bound optimizations for small input sizes typically underperform at larger input sizes.

\para{Demand for fine-grained communication.}
To minimize communication overhead, applications are now adopting new techniques to overlap communication with computation.
One such recent technique is \emph{distributed kernel fusion}, where communication and computation occur within the same GPU kernel, enabling even finer-grained overlap, such as at the granularity of tiles~\cite{10.1145/3767295.3769370}.
For example, FlashMoE overlaps tile communication with computation within a fused MoE kernel~\cite{NEURIPS2025_918d938b}, while MegaMoE pulls tokens over NVLink and computes within the same kernel~\cite{deepseekai2026}.
These applications need efficient, hardware-aware device-side primitives for expressing high-performance communication at the granularity of their computation.

\para{Impact on end-to-end performance.}
We argue that the demands discussed thus far are important because communication occupies a significant portion of end-to-end execution, specifically in LLM inference.
\Cref{fig:motivation} shows SGLang serving Qwen3.5-122B-A10B,
where communication accounts for 30--32\% of end-to-end latency.
Replacing its communication backend (predominantly NCCL) with \NAME reduces end-to-end communication time by up to 1.79$\times$, translating to up to 1.25$\times$ speedup in end-to-end latency.
We observe a similar gain in decode time (TPOT) and a 1.10$\times$ gain in prefill time (TTFT).

\subsection{Evolving hardware}\label{subsec:evolving-hardware}
\begin{table}[htbp]
  \centering
  \footnotesize
  \setlength{\tabcolsep}{5pt}
  \begin{tabular}{@{}lrrrr@{}}
    \toprule
    GPU & Bandwidth ($R$) & Latency ($\tau$) & $R\tau$ & Rounded BDP \\
    \midrule
    A100 & 300\,GB/s & 2.20\,\textmu s & 644\,KiB & 512\,KiB \\
    H200 & 450\,GB/s & 2.13\,\textmu s & 936\,KiB & 1\,MiB \\
    B200 & 900\,GB/s & 2.80\,\textmu s & 2.4\,MiB & 2\,MiB \\
    \bottomrule
  \end{tabular}
  \caption{BDP estimates using nominal unidirectional bandwidth and measured round-trip NVLink latency. The last column rounds to the closest power of 2. Latency results are obtained from \cref{fig:push-or-pull}.}
  \label{tab:background-bdp}
\end{table}
In an NVLink scale-up domain, GPU threads can read and write \emph{registered} memory from peer GPUs directly because of Unified Virtual Addressing (UVA)~\cite{nvidia2026cudaguide}.
As a result, frameworks implement communication in this domain via load and store instructions, like typical local memory accesses.
This datapath has evolved significantly across recent hardware generations: pre-Ampere GPUs relied on vectorized load and store instructions, until Ampere, which introduced \emph{asynchronous} loads from HBM to shared memory (on-chip) that avoid intermediate registers~\cite{nvidia2020ampere}.
Hopper added TMA for bulk, asynchronous transfers between HBM and shared memory, and NVLink SHARP for multicast stores and in-switch reduction~\cite{hopper_whitepaper}.
Meanwhile, inter-GPU interconnect bandwidth has increased exponentially from Ampere (300 GB/s unidirectional) to Blackwell (900 GB/s)~\cite{blackwell_tuning}.
Importantly, this rise in interconnect bandwidth requires communication frameworks to adopt newer data movement mechanisms to achieve peak performance.

\para{Faster interconnects require more bytes-in-flight.}
To sustain bandwidth $R$ over round-trip latency $\tau$, we must keep approximately the bandwidth-delay product (BDP)~\cite{bdp} in flight:
\begin{equation}\label{eq:background-bdp}
  Q \approx R\tau,
\end{equation}
where $Q$ is the number of outstanding bytes.
We measure round-trip latencies of NVLink on A100, H200 and B200 and observe approximately 2.2\,\textmu s, 2.1\,\textmu s and 2.8\,\textmu s, respectively.
Combined with increasing interconnect bandwidth, these imply a BDP growing from roughly 512\,KiB to 2\,MiB as shown in \cref{tab:background-bdp}.
To keep this much data in flight, an implementation can increase bytes-in-flight per Streaming Multiprocessor (SM) or allocate more SMs.
The latter trades off more hardware resources but is simpler to employ in practice and hence is commonly adopted in existing works.
The former is more efficient, demanding fewer resources, but requires (1) a \emph{capable} datapath and (2) tuning to achieve peak performance.
By ``capable,'' we mean a datapath equipped with mechanisms that allow for scaling the bytes-in-flight sustained \emph{within an SM} while maintaining efficient use of GPU resources, such as registers or shared memory.

\begin{figure}[t]
  \centering
  \includegraphics[width=\columnwidth]{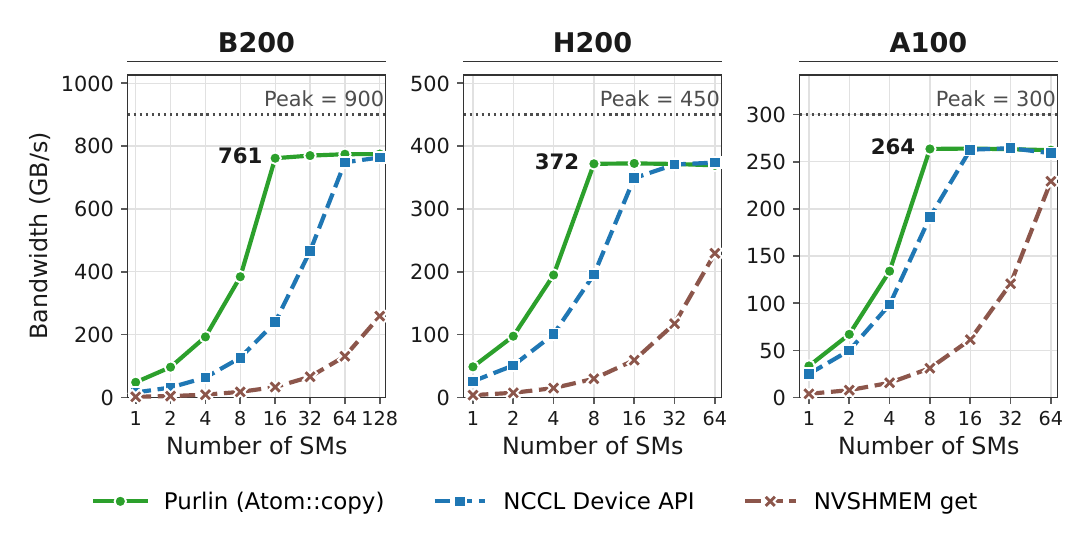}
  \caption{Bandwidth for a 256\,MiB pull over NVLink with 256 threads per SM. Results that are higher and to the left are better.}
  \label{fig:copy-microbench}
\end{figure}

\para{Capable, tunable datapaths use up to 4$\times$ fewer SMs.}
\Cref{fig:copy-microbench} compares three pull implementations over NVLink across three hardware generations.
\texttt{Atom::copy} uses a pipelining approach we explain in~\cref{subsubsec:unified-pipeline}, whereas the other systems use vectorized load and store instructions.
One system (NCCL Device API) allows for programmable tuning, whereas the other (NVSHMEM) does not.
We tune NCCL Device API by statically increasing its per-SM bytes-in-flight via loop unrolling to registers.
As we see, \texttt{Atom::copy}, the most capable datapath, occupies the Pareto frontier of bandwidth as a function of SM usage.
Specifically, NCCL's tuned device API approaches peak bandwidth at 16, 32, and 64 SMs on A100, H200, and B200, respectively; in contrast, \texttt{Atom::copy} reaches comparable or higher bandwidth with \emph{fewer} SMs at 8, 8, and 16.
On B200 and H200, this difference in SM usage is largest, where \texttt{Atom::copy} uses 4$\times$ fewer SMs to achieve peak measured bandwidth.
We also see that NVSHMEM, the least programmable, yields the lowest performance consistently across all three generations.
These results show that communication performance and resource efficiency strongly correlate with the capability and programmability of the underlying communication datapath, with performance highest for the most capable system, \texttt{Atom::copy}, and lowest for the least capable and non-tunable system, NVSHMEM.

\para{Requirements.}
We observe that ongoing evolution in hardware and interconnect bandwidth calls for \emph{adaptable} communication software (\cref{subsec:evolving-hardware}).
To achieve this adaptability, we argue that the software datapath must \emph{evolve} with hardware and that the resulting hardware-aware datapath should be accessible to upstream applications with fine-grained communication needs (\cref{subsec:evolving-applications}). Finally, we argue that the datapath should evolve \emph{separately} from orchestration, which also needs to adapt to application regimes (\cref{subsec:evolving-applications}).

Together, we arrive at these requirements:
\begin{enumerate}[label={\textbf{R\arabic*}},leftmargin=2.2em]
  \item \textbf{Separation of concerns:} separate orchestration from the datapath so both can evolve independently.
  \item \textbf{Regime specialization:} preserve specialization for application regimes within orchestration.
  \item \textbf{Programmable building blocks:} expose datapath primitives as building blocks for custom use cases.
\end{enumerate}

%% file: figures/purlin_breakdown_fig.tex
\begin{figure}[t]
  \centering
  \includegraphics[width=\columnwidth]{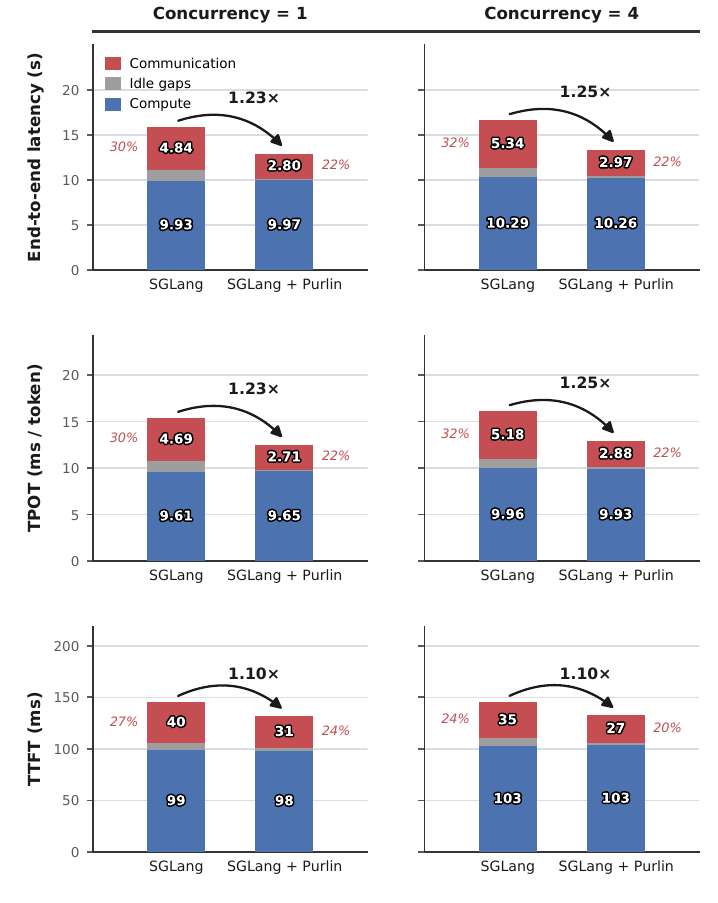}
  \caption{Latency breakdown for Qwen3.5-122B-A10B on eight A100 GPUs with TP8/EP8/DP4 and 1K input and 1K output tokens. Rows show p50 end-to-end latency, p50 Time Per Output Token (TPOT), and p50 Time To First Token (TTFT).}
  \label{fig:motivation}
\end{figure}

%% file: sections/3_design.tex
\section{Design}\label{sec:design}
\input{figures/design_listing_style}

We organize \NAME's design around the requirements outlined in \S\ref{sec:background}.
To decouple orchestration from the datapath of collectives (R1), we capture semantics through layouts (\S\ref{subsec:layouts}) and express collectives as transformations from an input layout to an output layout.
We use SNAC to (1) derive the orchestration for these transformations, with separate execution for latency and throughput (R2, \S\ref{subsec:snac}) and (2) delegate data movement to the Atom, whose unified pipeline and deterministic reduction we describe in \S\ref{subsec:atom}.
We use \texttt{allReduce} as a running example to show how these three layers interact.

\subsection{Layouts}\label{subsec:layouts}
A \emph{layout} describes how a rank interprets buffer regions relative to other participating ranks.
Concretely, layouts determine which region a participant contributes as input or receives as output in a collective.
For a group of $n$ participating ranks, we use three layouts:
\texttt{packed}, \texttt{scattered}, and \texttt{transposed}.

A \texttt{packed} buffer holds one contiguous contribution while a \texttt{scattered} buffer contains $n$ contiguous partitions indexed by rank.
Partition $k$ holds the contribution assigned to rank $k$ in input \texttt{scattered} buffers, but holds the result supplied by rank $k$ for output scattered buffers.

A \texttt{transposed} layout describes an exchange of partitions.
Specifically, destination rank $r$ receives partition $r$ from each source rank in that order.
The above layouts assume uniform partition sizes across ranks; we express non-uniform partition sizes with a \texttt{V} suffix to each layout.

\input{figures/design_layouts}
\input{figures/collective_namings}

\para{Collectives as layout transformations.}
We express a collective as a \emph{transformation between an input and an output layout}.
Consider the three ranks in \cref{fig:design-layouts}.
We view each input as \texttt{scattered}, with each containing contiguous partitions indexed by rank.
We see that \texttt{reduceScatter} \emph{reduces} this \texttt{scattered} input, producing one contiguous \texttt{packed} result.
Then, \texttt{allGather} \emph{copies} these packed results to every rank and yields a \texttt{scattered} output.
In addition, these two transformations together produce the result of executing an \texttt{allReduce} on the input \texttt{scattered} buffer.

Listing~\ref{lst:collective-namings} makes this description concrete.
Each declaration  specifies an input and an output layout, a \emph{consume operation}, an Atom (\cref{subsec:atom}) and a codesign policy (\cref{subsec:implementation-policies}).
The consume operation describes whether a SNAC transformation involves copies, such as for \texttt{allGather}, or reductions such as for \texttt{reduceScatter}.

\para{Composing transformations.}
We note that for \texttt{allReduce}, calling \texttt{reduceScatter} and \texttt{allGather} independently would be inefficient, requiring redundant memory operations and limiting optimizations such as operator overlapping.
Instead, we use \texttt{compose<F, G>} to express the execution of $F$ followed by $G$ as a single transformation.
As Listing~\ref{lst:collective-namings} shows, we capture \texttt{allReduce} as \texttt{compose<reduceScatter, allGather>}, a \texttt{scattered}$\rightarrow$\texttt{scattered} transformation.
In \cref{app:rooted-collectives}, we discuss extending composition to rooted collectives.

\subsection{Stage, Notify, And Consume (SNAC)}\label{subsec:snac}
SNAC is the orchestration protocol that executes a layout transformation across participating ranks.
Specifically, SNAC uses the input and output layouts to derive a communication schedule for the collective.
This schedule operates at the granularity of a \emph{chunk}, which is a subpartition of both the input and output buffers for a collective.
\input{figures/snac_execution_paths}

\noindent\textbf{Key operations.}
In SNAC's throughput path (\cref{fig:snac-throughput}), each rank first \emph{stages} its input contribution from \emph{application buffers} into registered, internal \emph{peer-accessible memory}.
This staging is a local \emph{copy} from these ordinary buffers, which need not be registered, to internal staging buffers which are registered.
When the application provides a peer-accessible buffer, SNAC omits staging (see \cref{subsec:microbenchmarks} for an evaluation of this scenario).

After completing staging, ranks \emph{notify} other participating ranks.
Here, ranks notify with \texttt{release} memory ordering~\cite{cppreference_memory_order}, ensuring that consuming ranks observe the notification \emph{after} the staging copies have completed.

After notification, each rank transitions to a consumer role.
Specifically, every rank waits to receive notifications (with \texttt{acquire} memory ordering) from every other rank and, upon receiving one, \emph{consumes} from the released staging buffers via a copy or reduction.

\para{Deriving orchestration from layouts.}
SNAC is a templated metaprogram whose input parameters include the layouts describing a collective.
With these layouts, SNAC determines how to enforce dependencies.
For example, for the \texttt{packed} input layout, each rank notifies every other rank of its entire input contribution.
However, with a scattered input, each rank selectively notifies in rank order so that rank $k$ expects a notification for partition $k$ from every other rank.

Likewise, SNAC employs the collective's output layout and consume operation to direct how ranks consume from other ranks.
For the \texttt{packed} output layout, SNAC directs ranks to reduce across all other ranks. In contrast, with the \texttt{scattered} or \texttt{transposed} output layouts, SNAC makes each rank copy other ranks' contributions and place that data in a rank-ordered layout in the output buffer.

SNAC adopts Cooperative Thread Array (CTA) specialization and chunking for \emph{overlapping} staging with consumption.
Specifically, for $N$ CTAs allocated to a SNAC transformation, SNAC splits $N$ into two sets of CTAs: one group stages exclusively while the other consumes from participating ranks.
\input{figures/snac_layout_branches}
Staging CTAs cooperate to stage a single data chunk $c$.
Each CTA stages by invoking \texttt{Atom::copy} on a predetermined slice of the chunk.
Upon completing the copy, each CTA increments an atomic counter, with \texttt{acq\_rel} memory ordering, that counts finished CTAs.
The last staging CTA to finish announces to all remote ranks (\texttt{allGather}) or a specific rank (\texttt{reduceScatter}) that $c$ is available for consumption as depicted in Lines \ref{line:snac-notify-begin} --~\ref{line:snac-notify-end} of Listing~\ref{lst:snac-layout-branches}.

Meanwhile, consuming CTAs on each rank \emph{wait} for the notification of a chunk before consumption, which in SNAC is a \emph{pull} from remote GPU memory on a rank to local memory of the calling rank (\cref{fig:snac-throughput}).
On receiving the notification, these CTAs invoke either \texttt{Atom::copy} or \texttt{Atom::reduce} as specified by the collective's \texttt{ConsumeOp} parameter.
As exemplified in Line~\ref{line:snac-reduce}, for \texttt{reduceScatter} specifically, consuming CTAs invoke \texttt{Atom::reduce} on chunk $c$ from all ranks.
Line~\ref{line:snac-gather} shows that for \texttt{allGather}, a CTA waits for $c$ from an assigned rank \texttt{peer}, after which the CTA copies with \texttt{Atom::copy}.
For that branch specifically, SNAC maps CTAs to ranks to ensure that $c$ is copied from all ranks.

\para{Extending orchestration for compositions.}
In executing \texttt{compose<F, G>}, SNAC further delegates an additional group of consuming CTAs for $G$.
Concretely, for \texttt{allReduce = compose<reduceScatter, allGather>}
there are three CTA groups: (1) staging and (2) reducing CTAs for \texttt{reduceScatter} and (3) gathering CTAs for \texttt{allGather}.
SNAC uses codesign policies (\cref{subsec:implementation-policies}) to determine these allocations.
More importantly, all three CTA groups execute at the granularity of a chunk; as a result, SNAC can pipeline staging with reduction and the final gather phase.

For \texttt{allReduce} in particular, the last reducing CTA to complete a chunk announces that chunk's readiness to all ranks.
Each gathering CTA within each rank then calls \texttt{Atom::copy} to copy its assigned peer's result into its output.

\input{algorithms/snac}
Algorithm~\ref{alg:allreduce} summarizes these roles on rank $r$.
We write $X_r[k,c]$ for input chunk $c$ of partition $k$.
We denote $S_r[k,c]$ as the chunk's staging destination and $Y_k[c]$ as the reduced result at rank $k$ of chunks across all ranks.
We express $O_r[k,c]$ as the chunk's final destination.

\begin{figure}[htbp]
  \centering
  \includegraphics[width=\columnwidth]{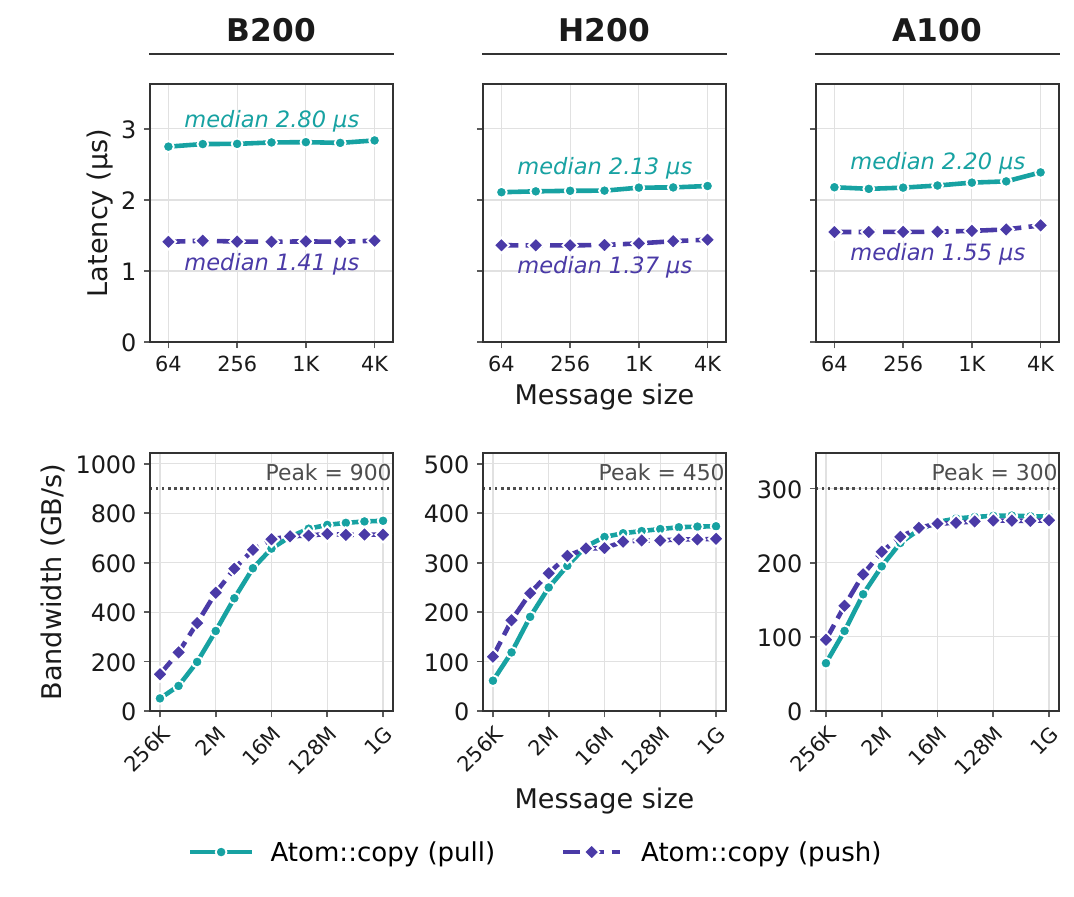}
  \caption{Push and pull over NVLink.
    We use 1 CTA for latency experiments and 16, 8, and 8 CTAs for bandwidth experiments on B200, H200, and A100, respectively. We use 256 threads per CTA for all experiments.}
  \label{fig:push-or-pull}
\end{figure}

\para{Specializing for latency.}
SNAC's execution as described thus far targets \emph{high throughput} specifically for medium to large messages which are bandwidth-bound rather than latency-bound.
Here, we explain a different execution path that SNAC adopts for latency-bound data sizes.

We first investigate empirically whether our use of remote reads for the throughput regime suits low-latency execution.
In \cref{fig:push-or-pull}, we compare pull (remote reads) with push (remote writes) through \texttt{Atom::copy} and observe up to 1.98$\times$ lower latency with push and up to 7\% higher bandwidth with pull.
This result confirms our choice of pull for throughput and guides our design choice of \emph{push for the latency regime}.

For latency-bound execution (\cref{fig:snac-latency}), SNAC combines staging and notification into a single memory operation.
Specifically, we package every 8 bytes of payload with an 8-byte flag (epoch in \cref{subsec:implementation-epochs}) into a 16-byte \emph{packet}, building on prior work~\cite{11244782, mscclpp}.
In our implementation, a producer rank atomically writes this packet into a consumer's peer-accessible buffer.
The consumer atomically polls for the expected flag and, after this condition holds, consumes the payload from that packet.
Unlike throughput-bound execution, which uses \texttt{release}–\texttt{acquire} ordering,
the latency path atomically transfers and observes payload and notification \emph{within} a single packet, permitting \texttt{relaxed}\footnote{We perform 16-byte \texttt{relaxed} atomic loads and stores via PTX instructions: \texttt{*.relaxed.sys.global.b128}, where * is \texttt{ld}~\cite{ptx_ld} or \texttt{st}~\cite{ptx_st}.} ordering.
As described, this technique \emph{doubles} the data volume transferred over the network and is therefore only performant in the latency-bound regime.
Hence, SNAC employs codesign policies (\cref{subsec:implementation-policies}) to determine the right crossover between using this path and the throughput path.
\subsection{The \texttt{Atom}}\label{subsec:atom}
\input{figures/atom_api}
As part of its orchestration, SNAC dispatches data movement to a supplied \texttt{Atom}, which provides efficient, hardware-aware copy and reduction operations for the executing GPU\@.
SNAC interacts with the \texttt{Atom} via the interface in Listing~\ref{lst:atom-api}.
\texttt{Atom::copy} transfers one contiguous buffer, while \texttt{Atom::reduce<E, op>} aggregates, via binary operation \texttt{op}, $n$ source buffers of type \texttt{E} into one destination.

\para{Architecture specialization.}
To maintain a unified Atom API to SNAC while preserving multiple implementations for different hardware generations, we leverage compile-time \emph{architecture specialization}.
This technique is inspired by existing device-side compute libraries cuBLASDx~\cite{nvidia2026cublasdx} and CUTLASS~\cite{nvidia_cutlass3}, which use it for exposing generation-specific ways to program tensor cores under a unified API.
Within the \ATOMTT, we realize this technique via the template parameter \texttt{GPUArch} which we use to represent a GPU generation.
For instance, as Listing~\ref{lst:atom_inst} shows, we identify Ampere (A100) with \texttt{GPUArch} set to \texttt{800} and implement \texttt{Atom<800>::copy} and \texttt{Atom<800>::reduce} by leveraging mechanisms specific to Ampere.
Moreover, this decoupled design \emph{localizes} hardware-specific details to the \texttt{Atom}, maintaining modularity within our stack.

\subsubsection{Unified Pipeline}\label{subsubsec:unified-pipeline}
Both \texttt{Atom::copy} and \texttt{Atom::reduce} use shared memory as workspace for a \emph{unified pipeline} which both operations lower to.
As motivated in \cref{subsec:evolving-hardware}, this pipeline enables the \texttt{Atom} to maximize concurrency \emph{within a CTA} allowing for higher achieved interconnect bandwidth with minimal SM usage.
\begin{figure*}[!t]
  \centering
  \captionsetup{skip=2pt}
  \input{figures/reduction_pipeline}
  \caption{Execution timeline of deterministic reduction via SCORE within \texttt{Atom::reduce}}
  \label{fig:deter_reduce}
  \vspace{-12pt}
\end{figure*}
\input{algorithms/atom_pipeline}

Algorithm~\ref{alg:atom-pipeline} shows the pipeline's three steps.
We first prime by asynchronously requesting data from global memory to shared memory (\texttt{workspace}) \emph{slots}, where each slot is a statically sized buffer dedicated to a pipeline stage.
We make $S$ such requests during priming, where $S$ is the number of pipeline stages.
In steady state, we wait for the oldest outstanding request to complete to its mapped slot.
Then, we consume from that slot into registers and asynchronously prefetch from global memory into that slot while processing the received data in registers.
We cycle across available slots when the total transfer stages $N$ exceed the total number of slots $S$.
We finally drain the pipeline over $S$ stages, where we process pending slots without any data requests.

\para{Programmability within the pipeline.}
We design the pipeline to be \emph{programmable}.
First, via the \texttt{Config} template parameter, we allow users to configure at compile-time parameters of the pipeline such as number of stages, number of participating threads and number of elements each thread processes.
We exemplify this config in \cref{fig:deter_reduce} and show an implementation in code in Listing~\ref{lst:atom_inst}.
Second, we allow users to configure the processing phase of the pipeline, allowing custom callbacks for operating on data while in registers.

\para{Asynchronous copy mechanism within the pipeline.}
For asynchronous copies from global to shared memory, on Ampere and above, we use the \texttt{cp.async} primitive~\cite{nvidia2020ampere}.
We retain \texttt{cp.async} on Hopper and Blackwell, rather than using TMA, because we observe 10--12\% higher NVLink bandwidth with \texttt{cp.async} compared to TMA. We believe this design decision also eliminates contention for the TMA unit with compute operations when both communication and computation are fused \emph{within the same SM}.
Specifically, \texttt{cp.async} uses the load/store unit (LSU), a distinct hardware resource from the TMA unit~\cite{nsight_compute_profiling}.
Typically, compute kernels use the TMA unit for operations like matrix multiplications; so, when both operations are fused within the same SM, our \texttt{Atom} would not contend for the TMA unit.
\input{figures/atom_instance}

\subsubsection{Deterministic Reduction for \texttt{Atom::reduce}}\label{subsubsec:score}
Machine learning frameworks increasingly demand deterministic floating-point reductions in collectives~\cite{sglang2025deterministic, he2025nondeterminism, mai_thinking_1} to ensure repeatable model outputs.
Floating-point reduction, such as addition, is not generally associative, meaning the result of a reduction \emph{depends on the order} in which the implementation combines inputs.
Existing implementations typically use optimizations (e.g., atomics or bespoke algorithms) that induce non-deterministic input orderings and thus produce non-deterministic outputs.
Removing such optimizations provides determinism but at the cost of performance.
We make progress in this tradeoff by equipping \texttt{Atom::reduce} with \textbf{S}equenced \textbf{Co}ncurrent \textbf{Re}duction (SCORE),\footnote{Like an orchestra, where multiple musical parts play concurrently, but the \emph{score} dictates their order.} which performs reductions strictly in \emph{source order} while achieving good performance via \emph{pipelining}.
In SCORE, each output element is
\begin{equation}
  y_i = \Bigl(\cdots\bigl((x_i^0 \oplus x_i^1) \oplus x_i^2\bigr) \oplus \cdots\Bigr) \oplus x_i^{n-1},
  \label{eq:score-order}
\end{equation}
where $x_i^s$ is element $i$ of source $s$ and $\oplus$ is the binary operation \texttt{op} in Listing~\ref{lst:atom-api}.
We convert each $x_i^s$ to FP32\footnote{For type E, we accumulate in \texttt{cuda::std::common\_type<float, E>}, which is \texttt{float} for fp16, bf16 and fp8 data types.}, accumulate in FP32, and convert only the final result back to the source datatype.
\cref{fig:deter_reduce} traces SCORE for addition over four sources on two threads.
We note that determinism within SCORE means that identical inputs produce bit-identical outputs across executions for a \emph{fixed} source ordering and hardware architecture.

We achieve determinism in SCORE via two invariants: \emph{exclusive ownership}, where exactly one thread computes an output element, and \emph{ordered accumulation}, namely, accumulations within a single thread occur in source order.
Exclusive ownership eliminates nondeterminism from multithreaded atomics.
Ordered accumulation expresses the constituent arithmetic for an output element in source order, relying on the \emph{sequenced-before}~\cite{cppreference_eval_order} relationship in the C++ execution model, in which we implement SCORE.

However, sequentially reading and reducing all $n$ sources would yield suboptimal performance, especially when sources are read from remote memory.
SCORE therefore runs on our unified pipeline (Algorithm~\ref{alg:atom-pipeline}) where a thread can keep multiple sources in flight while performing reductions.
Importantly, SCORE requests and processes data from the pipeline in source order (depicted in \cref{fig:deter_reduce}) to preserve the ordered accumulation invariant.
To increase throughput, each thread maintains multiple independent accumulators for distinct output elements (two in \cref{fig:deter_reduce}). This technique exploits Instruction-Level Parallelism (ILP) as updates to independent accumulators can execute concurrently in hardware.
SNAC performs all throughput-path reductions with \texttt{Atom::reduce}; therefore, every SNAC collective that uses SCORE is deterministic by construction by virtue of inheriting SCORE's determinism guarantee.
SNAC's latency path, which reduces packets, also follows both invariants: exactly one thread computes each output element and does so in source order.

\texttt{Atom::reduce} also provides an alternative reduction path via hardware-accelerated reduction in NVLink SHARP (NVLS)~\cite{9895480}, available on Hopper GPUs and above.
However, NVLS reduction, to the best of our knowledge, occurs in an unspecified order and does not provide the determinism guarantee of SCORE.
We therefore allow users to choose either SCORE or NVLS in \NAME.
We compare both implementations in \cref{app:multimem}.

\subsubsection{Pipeline Codesign Toward the BDP}\label{subsubsec:pipe-codesign}
To achieve high bandwidth with the pipeline, we need to tune its capacity to maximize bytes-in-flight.
We use the BDP in \cref{eq:background-bdp} as our maximal target for bytes-in-flight.
Precisely, if each of $C$ CTAs has $S$ pipeline stages, each holding $B_{\mathrm{stage}}$ bytes, the pipeline capacity or maximal bytes-in-flight sustained by the GPU is
\begin{equation}
  Q_{\mathrm{cap}} = C S B_{\mathrm{stage}}.
  \label{eq:pipeline-window}
\end{equation}
Thus, the \textbf{goal} of our tuning is to get $Q_{\mathrm{cap}}$ to approach the BDP while minimizing $C$.
To get there, we can increase $B_{\mathrm{stage}}$ or $S$, which increase concurrency within a CTA but consume proportionally higher shared memory on the hosting SM.
Increasing $C$, by contrast, allocates more CTAs, increasing the number of SMs used.
We tune to find the best values for $C$, $S$, and $B_{\mathrm{stage}}$ and codify those choices in codesign policies (\cref{subsec:implementation-policies}).
Moreover, we empirically isolate the effect of $S$ on bandwidth performance at fixed $C$ via an ablation in \cref{subsubsec:pipeline-ablation}.

%% file: figures/design_listing_style.tex
\setminted[cpp]{
  fontencoding=T1,
  fontfamily=txtt,
  fontsize=\footnotesize,
  bgcolor=white,
  frame=single,
  framerule=0.4pt,
  framesep=4pt,
  linenos,
  numbersep=6pt,
  xleftmargin=14pt,
  breaklines,
  tabsize=2
}

%% file: figures/design_layouts.tex
\begin{figure*}[t]
  \centering
  \resizebox{\textwidth}{!}{%
  \begin{tikzpicture}[
    x=1cm,y=1cm,
    font=\small,
    cell/.style={draw=black!65,minimum width=0.94cm,minimum height=0.48cm,inner sep=1pt},
    flow/.style={-{Latex[length=2.2mm]},thick},
    rank/.style={anchor=east,font=\small}
  ]
    \node at (2.5,0.65) {Input: \texttt{scattered}};
    \node at (8.3,0.65) {Intermediate: \texttt{packed}};
    \node at (14.3,0.65) {Output: \texttt{scattered}};
    \foreach \p in {0,1,2} {
      \node[rank] at (0.83,-0.62*\p) {rank \p};
      \node[rank] at (7.6,-0.62*\p) {rank \p};
      \node[rank] at (12.63,-0.62*\p) {rank \p};
      \foreach \k/\shade in {0/blue!12,1/teal!17,2/orange!20} {
        \node[cell,fill=\shade] at (1.5+\k,-0.62*\p) {$x_{\p,\k}$};
        \node[cell,fill=\shade] at (13.3+\k,-0.62*\p) {$y_{\k}$};
      }
    }
    \foreach \k/\shade in {0/blue!12,1/teal!17,2/orange!20} {
      \node[cell,fill=\shade] at (8.3,-0.62*\k) {$y_{\k}$};
    }
    \draw[flow] (4.15,-0.62) -- (6.35,-0.62)
      node[midway,above=4pt] {ReduceScatter};
    \node[align=center,font=\small] at (5.4,-1.3) {reduce each\\input column};
    \draw[flow] (9.4,-0.62) -- (11.5,-0.62)
      node[midway,above=4pt] {AllGather};
    \draw[black!25] (0,-2.0) -- (15.9,-2.0);
    \node[anchor=west] at (0,-2.72) {AllToAll at rank 1:};
    \node at (6.25,-2.35) {Input: \texttt{scattered}};
    \node at (13.5,-2.35) {Output: \texttt{transposed}};
    \foreach \k/\shade in {0/blue!12,1/teal!17,2/orange!20} {
      \node[cell,fill=\shade] at (5.25+\k,-2.85) {$x_{1,\k}$};
      \node[cell,fill=teal!17] at (12.5+\k,-2.85) {$x_{\k,1}$};
    }
    \draw[flow] (8.1,-2.85) -- (11.55,-2.85)
      node[midway,above=3pt] {exchange partitions};
  \end{tikzpicture}%
  }
  \caption{Layouts and collectives as layout transformations. Intermediate serves as both an output and input layout.}
  \label{fig:design-layouts}
\end{figure*}

%% file: figures/collective_namings.tex
\begin{listing*}[t]
\begin{minted}{cpp}
using enum ConsumeOp; // {copy, reduce}
using enum DataLayout; // {packed(V), scattered(V), transposed(V)}

using reduceScatter  = SNAC<Atom, Policy, reduce, scattered,  packed>;
using reduceScatterV = SNAC<Atom, Policy, reduce, scatteredV, packedV>;
using allGather      = SNAC<Atom, Policy, copy,   packed,     scattered>;
using allGatherV     = SNAC<Atom, Policy, copy,   packedV,    scatteredV>;
using allToAll       = SNAC<Atom, Policy, copy,   scattered,  transposed>;
using allToAllV      = SNAC<Atom, Policy, copy,   scatteredV, transposedV>;

// (scattered -> packed) o (packed -> scattered), in execution order.
using allReduce      = compose<reduceScatter, allGather>; // scattered -> scattered
\end{minted}
\caption{Collective declarations select an Atom, a policy, a consume operation, and a pair of input and output layouts.}
\label{lst:collective-namings}
\end{listing*}

%% file: figures/snac_execution_paths.tex
\begin{figure*}[!t]
  \centering
  \captionsetup{skip=3pt}
  \captionsetup[subfigure]{skip=1pt}
  \begin{subfigure}[b]{0.476\textwidth}
    \centering
    \includegraphics[width=\linewidth, trim=17.5 17.5 17.5 17.5, clip]{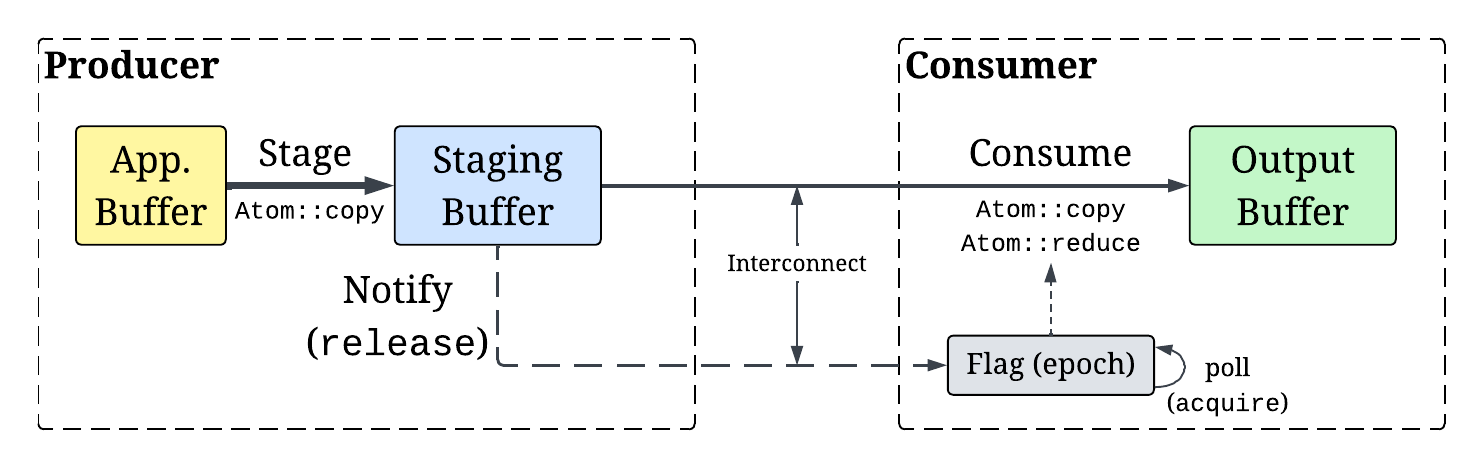}
    \caption{Throughput-bound execution.}
    \label{fig:snac-throughput}
  \end{subfigure}\hfill
  \begin{subfigure}[b]{0.508\textwidth}
    \centering
    \includegraphics[width=\linewidth, trim=17.5 17.5 17.5 17.5, clip]{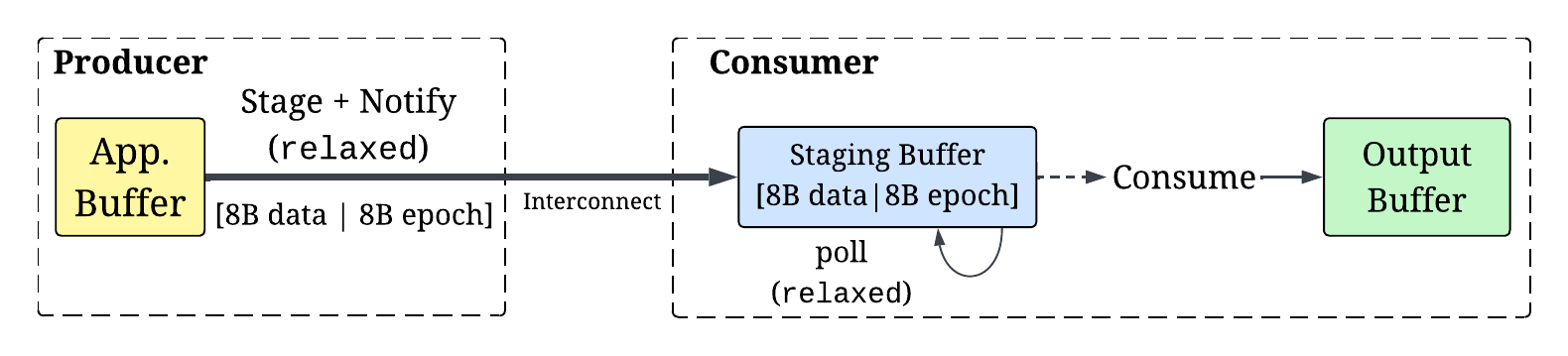}
    \caption{Latency-bound execution.}
    \label{fig:snac-latency}
  \end{subfigure}
  \caption{Overview of SNAC.}
  \label{fig:snac-execution-paths}
\end{figure*}

%% file: figures/snac_layout_branches.tex
\begin{listing}[!ht]
\begin{minted}[escapeinside=||]{cpp}
// Stage then last staging CTA announces chunk c.
if constexpr (InputLayout == packed)|\phantomsection\label{line:snac-notify-begin}|
  notifyAll(c);
else
  // notify rank k only
  notifyOne(k, c);|\phantomsection\label{line:snac-notify-end}|

if constexpr (Output == packed)
  out = dst + offset;
else
  out = dst + peer * partBytes + offset;

if constexpr (consumeOp == reduce) {
  wait_all_ready(c);
  Atom::reduce<E>(out, sources, bytes, n, workspace); |\phantomsection\label{line:snac-reduce}|
} else {
  wait_ready(peer, c); |\phantomsection\label{line:snac-gather}|
  Atom::copy(out, src, bytes, workspace);
}
\end{minted}
\caption{Selected rules in SNAC.}
\label{lst:snac-layout-branches}
\end{listing}

%% file: algorithms/snac.tex
\begin{algorithm}[t]
  \caption{Composed AllReduce at rank $r$}
  \label{alg:allreduce}
  \small
  \begin{algorithmic}[1]
    \Require $n$ ranks; $b$ slice of a chunk; $s$ staging slots; workspace $w$
    \Statex \emph{Run the three role loops concurrently.}
    \Statex \textbf{ReduceScatter: staging CTAs}
      \For{each assigned input partition $k$ and chunk $c$}
        \State \textbf{if} $c \ge s$: await $\mathrm{reduced}(k, c{-}s)$ \textbf{if} $k \ne r$,
        \Statex \hspace{\algorithmicindent}\hspace{\algorithmicindent}\textbf{else} await $\mathrm{consumed}(p, r, c{-}s)$ for all ranks $p$
        \State $\texttt{Atom::copy}(S_r[k,c],X_r[k,c],b,w)$
        \State Last producer publishes $\operatorname{staged}(r,k,c)$ to rank $k$
      \EndFor
    \Statex \textbf{ReduceScatter: reducer CTAs}
      \For{each chunk $c$ of partition $r$}
        \State Await $\operatorname{staged}(p,r,c)$ for every rank $p$
        \State $P\gets[S_0[r,c],\ldots,S_{n-1}[r,c]]$
        \State $\texttt{Atom::reduce<E>}(Y_r[c],P,b,n,w)$
        \State Last reducer publishes $\operatorname{reduced}(r,c)$ to all ranks
      \EndFor
    \Statex \textbf{AllGather: gather CTAs (consume only)}
      \For{each assigned result partition $k$ and chunk $c$}
        \State Await $\operatorname{reduced}(k,c)$
        \State $\texttt{Atom::copy}(O_r[k,c],Y_k[c],b,w)$
        \State Last gather CTA publishes consumed$(r,k,c)$ to rank $k$.
      \EndFor
  \end{algorithmic}
\end{algorithm}

%% file: figures/atom_api.tex
\begin{listing}[!h]
\begin{minted}{cpp}
template<int GPUArch, typename Config>
struct Atom {
  static void copy(
    std::byte* dst,
    std::byte* src,
    size_t bytes,
    std::byte* workspace);

  template<typename E, RedOp op = Op::add>
  static void reduce(
    std::byte* dst,
    std::byte** sources,
    size_t bytes,
    int n,
    std::byte* workspace);
};
\end{minted}
\caption{Atom's interface.}
\label{lst:atom-api}
\vspace{-12pt}
\end{listing}

%% file: figures/reduction_pipeline.tex
\resizebox{\textwidth}{!}{%
\begin{tikzpicture}[
  x=1pt, y=-1pt,
  font=\fontsize{8}{9}\selectfont,
  score box/.style={draw=black!80, line width=0.75pt, rounded corners=1pt, inner sep=0pt},
  score fetch/.style={score box, minimum width=66pt, minimum height=16pt},
  score process/.style={score fetch, fill=white},
  score result/.style={score fetch, fill=scoreResult, minimum height=14pt},
  score divider/.style={draw=black!60, densely dotted, line width=0.55pt},
  score flow/.style={draw=black!80, line width=0.75pt, -{Stealth[length=3.5pt,width=2.8pt]}},
  score span/.style={draw=black!80, line width=0.75pt, {Stealth[length=3.5pt,width=2.8pt]}-{Stealth[length=3.5pt,width=2.8pt]}},
  score math/.style={font=\fontsize{9}{10}\selectfont, inner sep=0pt},
  score update/.style={font=\fontsize{7.5}{8.5}\selectfont, align=center, inner sep=0pt},
  score label/.style={font=\fontsize{7}{8}\selectfont, inner sep=0pt}
]
  \definecolor{scoreSourceZero}{HTML}{F5D6ED}
  \definecolor{scoreSourceOne}{HTML}{FFF5AE}
  \definecolor{scoreSourceTwo}{HTML}{E3E1C8}
  \definecolor{scoreSourceThree}{HTML}{B9EEEA}
  \definecolor{scoreResult}{HTML}{C7F0CB}

  \foreach \source/\sourceColor in {0/scoreSourceZero,1/scoreSourceOne,2/scoreSourceTwo,3/scoreSourceThree} {
    \pgfmathsetmacro{\sourceCenter}{30+16*\source}
    \node[anchor=east, inner sep=0pt] at (32,\sourceCenter) {\textbf{source}~\source};
    \foreach \element in {0,1,2,3} {
      \pgfmathsetmacro{\elementCenter}{44+16*\element}
      \node[score box, fill=\sourceColor, minimum width=16pt, minimum height=16pt]
        at (\elementCenter,\sourceCenter) {};
      \node[score math] at (\elementCenter,\sourceCenter) {$x_{\element}^{\source}$};
    }
  }

  \node[score box, minimum width=64pt, minimum height=16pt] (reduce) at (68,110)
    {\texttt{Atom::reduce}};
  \node[anchor=east, inner sep=0pt] at (32,134) {\textbf{Dest.}};
  \foreach \element in {0,1,2,3} {
    \pgfmathsetmacro{\elementCenter}{44+16*\element}
    \draw[score flow] (\elementCenter,86) -- (\elementCenter,102);
    \draw[score flow] (\elementCenter,118) -- (\elementCenter,126);
    \node[score box, fill=scoreResult, minimum width=16pt, minimum height=16pt]
      at (\elementCenter,134) {};
    \node[score math] at (\elementCenter,134) {$y_{\element}$};
  }
  \node[score label] at (68,154) {where $y_i=\displaystyle\sum_{s=0}^{3}x_i^s$};
  \draw[black!65, line width=0.65pt] (113,9) -- (113,166);

  \node[inner sep=0pt] at (316,4)
    {\textbf{Pipeline Config:} \{Threads per CTA = 2, Stages = 2, Elements per stage = 2\}};

  \foreach \thread/\firstElement/\secondElement/\fetchCenter/\processCenter in {0/0/2/30/55,1/1/3/106/131} {
    \node[anchor=east, font=\fontsize{7.5}{8.5}\selectfont, inner sep=0pt]
      at (150,\fetchCenter) {\textbf{thread \thread}};
    \node[score update] at (191,\processCenter)
      {$a_{\firstElement}=0$, $a_{\secondElement}=0$\\accumulators};

    \foreach \source/\sourceColor in {0/scoreSourceZero,1/scoreSourceOne,2/scoreSourceTwo,3/scoreSourceThree} {
      \pgfmathsetmacro{\fetchPosition}{191+72*\source}
      \pgfmathsetmacro{\processPosition}{263+72*\source}
      \node[score fetch, fill=\sourceColor] (fetch-\thread-\source)
        at (\fetchPosition,\fetchCenter) {};
      \draw[score divider]
        (\fetchPosition,{\fetchCenter-8}) -- (\fetchPosition,{\fetchCenter+8});
      \node[score math] at ({\fetchPosition-16.5},\fetchCenter) {$x_{\firstElement}^{\source}$};
      \node[score math] at ({\fetchPosition+16.5},\fetchCenter) {$x_{\secondElement}^{\source}$};

      \node[score process] (process-\thread-\source)
        at (\processPosition,\processCenter) {};
      \draw[score divider]
        (\processPosition,{\processCenter-8}) -- (\processPosition,{\processCenter+8});
      \foreach \element/\offset in {\firstElement/-16.5,\secondElement/16.5} {
        \node[score update] at ({\processPosition+\offset},\processCenter)
          {$a_{\element}\mathrel{+{=}}x_{\element}^{\source}$};
      }
      \draw[score flow] (fetch-\thread-\source.south east)
        to[out=-65,in=115] (process-\thread-\source.north west);
    }

    \foreach \source/\nextSource in {0/2,1/3} {
      \draw[score flow, densely dashed] (process-\thread-\source.north east)
        to[out=65,in=-115] (fetch-\thread-\nextSource.south west);
    }

    \node[score result] (result-\thread) at (479,{\processCenter+28}) {};
    \draw[score divider] (479,{\processCenter+21}) -- (479,{\processCenter+35});
    \draw[score flow] (process-\thread-3.south) -- (result-\thread.north);
    \foreach \element/\offset in {\firstElement/-16.5,\secondElement/16.5} {
      \node[score update] at ({479+\offset},{\processCenter+28}) {$y_{\element}=a_{\element}$};
    }
  }

  \draw[score span] (191,22) -- (191,17) -- (263,17) -- (263,22);
  \node[score label] at (227,12) {prime};
  \node[score label, anchor=east] at (224,43) {process};
  \node[score label, anchor=west] at (309,42) {prefetch};

  \draw[score span] (263,63) -- (263,71) -- (335,71) -- (335,63);
  \node[score label] at (299,77) {steady state};
  \draw[score span] (407,63) -- (407,71) -- (463,71) -- (463,63);
  \node[score label] at (435,77) {drain};
  \draw[score flow] (365,159) -- (419,159);
  \node[score label, anchor=west] at (423,159) {time};
\end{tikzpicture}%
}%

%% file: algorithms/atom_pipeline.tex
\begin{algorithm}[t]
  \caption{Unified pipeline}
  \label{alg:atom-pipeline}
  \small
  \begin{algorithmic}[1]
    \Require $N$ total transfer stages, $S$ pipeline stages, $N\geq S$
    \For{$j=0,\ldots,S-1$} \Comment{prime}
      \State Issue asynchronous request of stage $j$ into slot $j$
    \EndFor
    \For{$j=S,\ldots,N-1$} \Comment{steady state}
      \State $s\gets j\bmod S$
      \State Wait for stage $j-S$ in slot $s$
      \State Read slot $s$ into registers $v$
      \State Issue asynchronous request of stage $j$ into slot $s$
      \State Process $v$: store a copy or update the running reduction
    \EndFor
    \For{$j=N-S,\ldots,N-1$} \Comment{drain}
      \State Wait for stage $j$ in slot $j\bmod S$
      \State Read the slot into registers $v$
      \State Process $v$: store a copy or update the running reduction
    \EndFor
  \end{algorithmic}
\end{algorithm}

%% file: figures/atom_instance.tex
\begin{listing}[t]
  \captionsetup{skip=0pt, belowskip=0pt}
\begin{minted}{cpp}
template<int GPUArch>
struct Config {
  static constexpr int THREADS = 256; // per CTA
  static constexpr int PIPELINE_STAGES = 8;
  static constexpr int ELEMS_PER_STAGE =
    GPUArch >= 900 ? 4 : 2;
  static constexpr int BYTES_PER_ELEM = 16;
  static constexpr int STAGE_BYTES =
    THREADS * ELEMS_PER_STAGE * BYTES_PER_ELEM;
  static constexpr int UNROLL = 8;
};
// sample Atoms
using GenericAtom   = Atom<700, Config<700>>;
using AmpereAtom    = Atom<800, Config<800>>;
using HopperAtom    = Atom<900, Config<900>>;
using BlackwellAtom = Atom<1000, Config<1000>>;
\end{minted}
\vspace{-10pt}\caption{Instantiations of the Atom.}
\label{lst:atom_inst}
\end{listing}

%% file: sections/4_implementation.tex
\section{Implementation}\label{sec:implementation}
We implement \NAME as a CUDA C++20 header-only library and also provide a Python API.
\NAME comprises 6,200 lines of CUDA C++ (\cref{tab:implementation-loc}) and 704 lines of Python.
We build \NAME from scratch, depending only on the CUDA C++ standard library libcu++~\cite{nvidia_libcudacxx} and CUB~\cite{nvidia_cub}.
\begin{table}[!h]
    \centering
    \small
    \begin{tabular}{@{}lr@{}}
        \toprule
        Component & Lines of code \\
        \midrule
        Generic Atom & 189 \\
        Ampere Atom & 300 \\
        Hopper Atom & 296 \\
        Blackwell Atom & 52 \\
        SNAC & 1,984 \\
        Codesign policies & 628 \\
        \midrule
        Utilities & 2,751 \\
        \textbf{Total} & 6,200 \\
        \bottomrule
    \end{tabular}
    \caption{Lines of CUDA C++ in \NAME}
    \label{tab:implementation-loc}
    \vspace{-12pt}
\end{table}

\para{Initialization and Python API.}
At initialization, \NAME's C++ API accepts pointers to registered buffers which we
use as staging buffers (\cref{subsec:epoch-staging}).
For the Python API, we use PyTorch symmetric memory~\cite{wang2025symmetricmemory} to allocate these buffers.
In Python, we also do JIT compilation where we supply an architecture identifier (\texttt{GPUArch} in Listing \ref{lst:atom-api})
to instantiate an Atom implementation and select codesign policies specific to the supplied GPU architecture\@.
Python calls pass ordinary PyTorch tensors and a CUDA stream to this compiled module.
\subsection{Staging Buffers}\label{subsec:epoch-staging}
Per GPU, we allocate two 8 MiB staging buffers for latency-bound execution and two 256\,MiB staging buffers for throughput-bound execution.
Across SNAC calls, we alternate between buffers as an optimization to sidestep inter-GPU synchronization within a call.
Users can configure the capacity of these buffers at initialization, trading memory use for performance.
When a collective's staging footprint exceeds the staging buffer size in the throughput regime, we cycle through a fixed number of slots which is $\frac{stagingSize}{chunkSize}$,
where $chunkSize$ is specified by the codesign policy passed to SNAC.
\input{figures/codesign_policy}
\subsection{Codesign Policies and Tuning}\label{subsec:implementation-policies}
Codesign policies capture parameters that configure SNAC's execution.
Example parameters are the latency--throughput crossover, chunk sizes, CTA allocations, and \texttt{Atom} pipeline stages (\cref{subsec:atom}).
We tune these parameters offline for each collective and GPU generation and codify our findings as
\emph{compile-time constants} which our APIs use.
Applications can also define their own policies and configure SNAC as needed.
We show a simplified policy in Listing~\ref{lst:codesign-policy}.
\subsection{Epochs}\label{subsec:implementation-epochs}
To reuse staging buffers safely, we use \emph{epochs}, which are monotonic 64-bit integers independently maintained by each CTA on a GPU.
Specifically, we use epoch parity to select one of the two internal staging buffers.
This epoch also serves as the value used for notification in SNAC for both throughput- and latency-bound execution (\cref{fig:snac-execution-paths}).
For epoch increments, we use the number of chunks in the current SNAC collective, rounded to an odd number
so consecutive calls safely alternate buffers.
For variable-length collectives with varying chunk counts, we use the maximum across all ranks.

%% file: figures/codesign_policy.tex
\begin{listing}[!h]
\begin{minted}{cpp}
using enum StagingMode; // {cyclic, resident}
using enum Regime; // {latency, throughput}
struct Policy {
  static constexpr size_t CHUNK_SIZE = 1024 * 1024;
  static constexpr int STAGE_BLOCKS = 16;
  static constexpr auto STAGE_MODE = cyclic;
  static constexpr auto REGIME = throughput;
};

// Within SNAC(args)
if constexpr (Policy::REGIME == latency) {
  runPackets(args);
}
else {
  // throughput regime
  if (blockIdx.x < Policy::STAGE_BLOCKS)
    stage<Policy>(args);
  else
    consume<Policy>(args);
}
\end{minted}
\caption{A simplified codesign policy and its usage in SNAC.}
\label{lst:codesign-policy}
\vspace{-12pt}
\end{listing}

%% file: sections/5_evaluation.tex
\section{Evaluation}\label{sec:evaluation}

Our evaluation addresses three questions:
\begin{enumerate}[label=\textbf{Q\arabic*.},leftmargin=2.2em,itemsep=2pt,topsep=3pt,parsep=0pt]
  \item Can collectives derived from SNAC deliver competitive performance across multiple GPU generations and latency and throughput regimes? (\cref{subsec:microbenchmarks})
  \item How do SNAC, the \ATOMTT and codesign policies independently contribute to \NAME's performance? (\cref{subsec:ablation})
  \item When do inference applications benefit, and how does \NAME affect output quality? (\cref{subsec:application-performance} and \cref{subsec:qualitative-evaluation})
\end{enumerate}

\begin{figure*}[!t]
  \centering
  \begin{subfigure}[t]{0.495\textwidth}
    \centering
    \includegraphics[width=\linewidth]{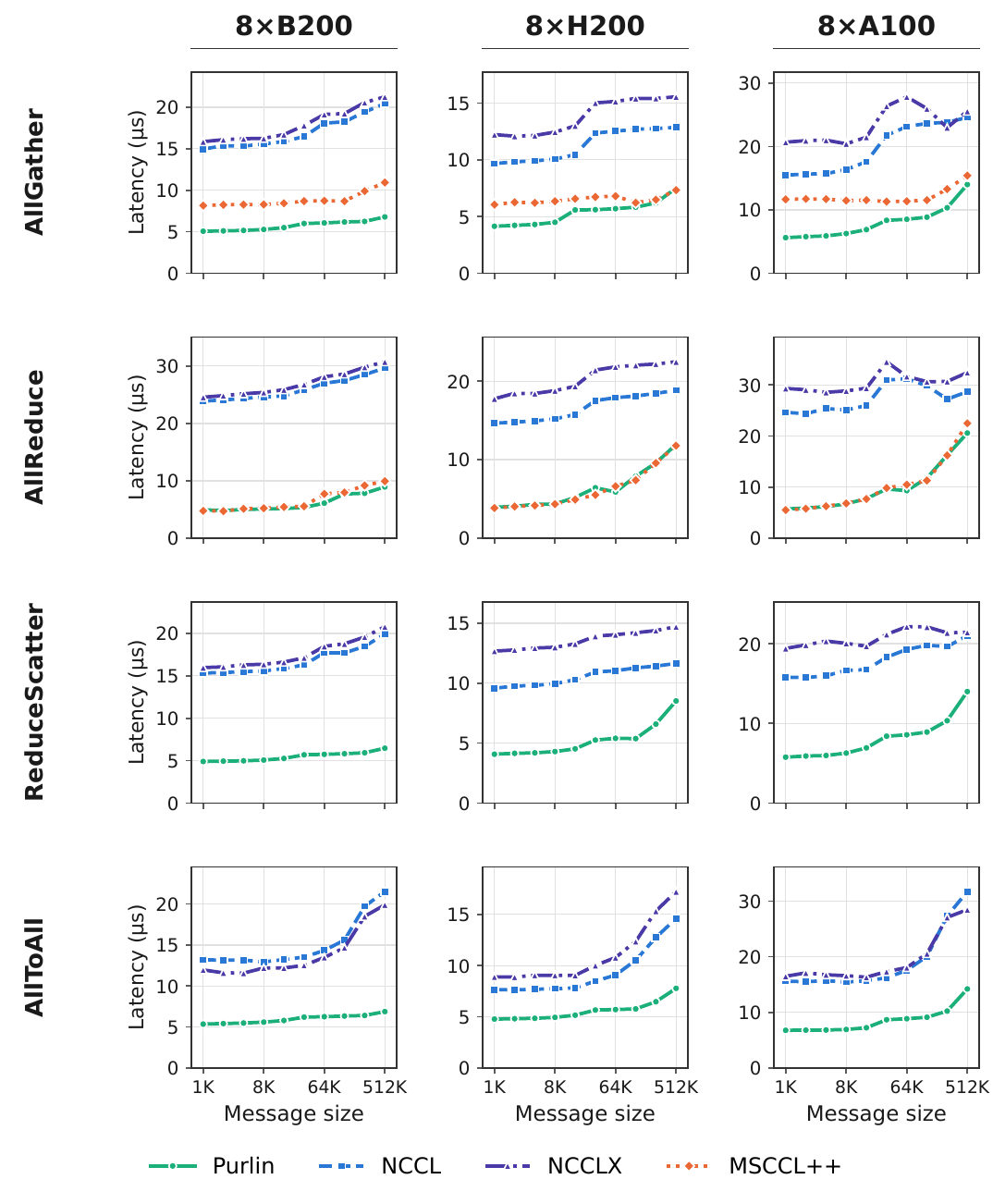}
    \caption{Latency. Lower is better.}
    \label{fig:eval-collective-latency}
  \end{subfigure}\hfill
  \begin{subfigure}[t]{0.495\textwidth}
    \centering
    \includegraphics[width=\linewidth]{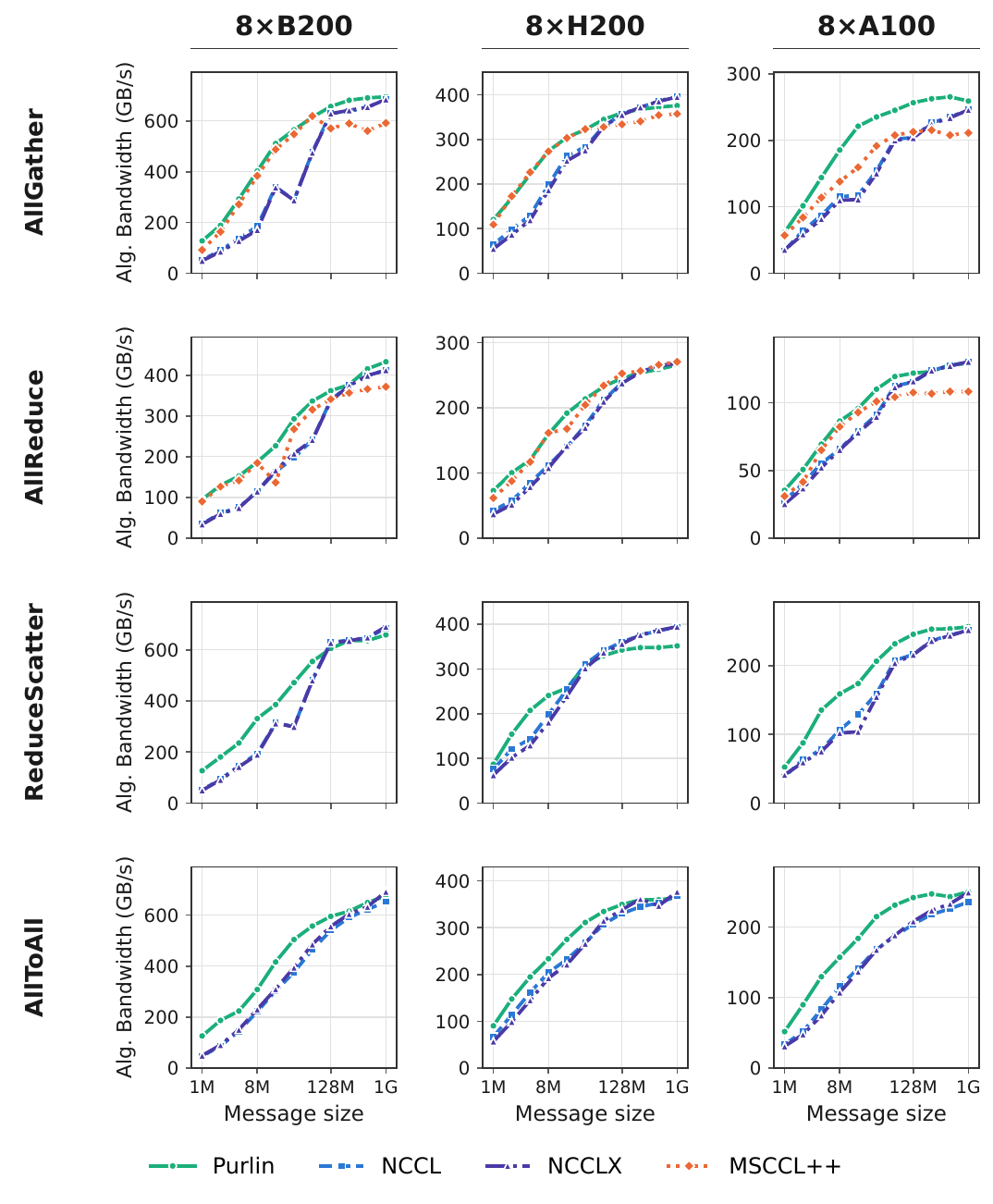}
    \caption{Algorithm bandwidth. Higher is better.}
    \label{fig:eval-collective-bandwidth}
  \end{subfigure}
  \caption{Collective communication performance on ordinary buffers.}
  \vspace{-12pt}
  \label{fig:eval-collectives}
\end{figure*}

\para{Experimental setup.}
We use three servers spanning three GPU generations, each with eight GPUs connected by NVLink (\cref{tab:eval-platforms}).
\begin{table}[ht]
  \centering
  \small
  \begin{tabular}{@{}llr@{}}
    \toprule
    \shortstack[l]{GPU and\\interconnect} & \shortstack[l]{LLM checkpoint\\(precision)} & \shortstack[r]{Total\\parameters} \\
    \midrule
    \begin{tabular}[c]{@{}l@{}}A100-SXM4 (80 GB)\\NVLink 3\end{tabular} & \begin{tabular}[c]{@{}l@{}}Qwen3.5-122B-A10B~\cite{qwen35_model}\\(BF16)\end{tabular} & 122B \\
    \addlinespace[3pt]
    \begin{tabular}[c]{@{}l@{}}H200 (141 GB)\\NVLink 4\end{tabular} & \begin{tabular}[c]{@{}l@{}}DeepSeek-V4-Flash~\cite{sglproject2026dsv4flashfp8}\\(FP8)\end{tabular} & 291B \\
    \addlinespace[3pt]
    \begin{tabular}[c]{@{}l@{}}B200 (180 GB)\\NVLink 5\end{tabular} & \begin{tabular}[c]{@{}l@{}}DeepSeek-V4-Pro~\cite{nvidia2026dsv4pronvfp4}\\(NVFP4)\end{tabular} & 1.6T \\
    \bottomrule
  \end{tabular}
  \caption{Evaluation platforms and LLM configurations.}
  \label{tab:eval-platforms}
\end{table}

\para{Baselines.}
We compare \NAME with established collective communication libraries and specialized GPU communication kernels.
NCCL 2.29.7~\cite{nvidia2025nccl} is NVIDIA's general-purpose collective communication library.
NCCLX~\cite{ncclx}, accessed through TorchComms 0.3.0~\cite{torchcomms2025}, is Meta's extension of NCCL.
We evaluate both on AllGather, ReduceScatter, AllToAll, AllReduce, and the three variable-length collectives.
For NCCL AllToAll, we use \texttt{ncclSend} and \texttt{ncclRecv} wrapped with \texttt{ncclGroupStart} and \texttt{ncclGroupEnd}.

In addition, we compare with MSCCL++ 0.10.0~\cite{mscclpp} which provides only AllGather and AllReduce, both as manually developed implementations.
Also, we compare with ParallelKittens~\cite{parallelkittens} (commit \texttt{67845f5}) which provides specialized kernels for Hopper and Blackwell.
For \cref{fig:copy-microbench}, we use NVSHMEM 3.7.2~\cite{nvidia2026nvshmem}.

\para{End-to-end inference integration.}
We integrate \NAME as a communication backend into SGLang~\cite{sglang} at revision \texttt{0f18d38} for end-to-end experiments on the models in \cref{tab:eval-platforms}.
For baseline SGLang, its communication backend is NCCL. We also integrate NCCLX into SGLang and extend SGLang's existing integration of MSCCL++.
We compare these systems in offline LLM serving, online LLM serving on a real-world production trace~\cite{mooncake_trace}, and diffusion image generation.
Offline serving includes all three integrations while online serving includes \NAME and MSCCL++.
We omit NCCLX from online serving because we experienced memory stranding issues when tearing down its process groups.
For image generation, we do not evaluate MSCCL++ because MSCCL++ does not provide an AllToAll implementation.

\para{Microbenchmark buffer configurations.}
We evaluate microbenchmarks under two application buffer configurations: (1) \textbf{ordinary} buffers which are allocated via \texttt{cudaMalloc} and are not registered with CUDA IPC~\cite{nvidia2026cudaguide} and (2) \textbf{peer-accessible} buffers which are registered.
For peer-accessible buffers, we use NCCL symmetric~\cite{ncclsymm} which we label as NCCL Symm.
In our experiments for NCCL Symm, we allocate input and output buffers via \texttt{ncclMemAlloc} and register using \texttt{ncclCommWindowRegister} with the flag \texttt{NCCL\_WIN\_COLL\_SYMMETRIC}.

For MSCCL++ AllReduce on B200 with ordinary buffers, we ensure a fair comparison by modifying the algorithm selector of MSCCL++ to pick non-zero-copy algorithms from its tuned candidates.
Otherwise, we use MSCCL++ code unmodified and also set \texttt{MSCCLPP\_NCCL\_SYMMETRIC\_MEMORY=1} for peer-accessible buffers.

We evaluate ParallelKittens with only peer-accessible buffers because that is the only configuration they support.
We evaluate NCCLX for ordinary buffers alone because, to the best of our knowledge, NCCLX does not provide API configurations for peer-accessible buffers.
We evaluate \NAME for peer-accessible buffers as \NAME-ZS, which omits staging within SNAC because the input buffers are already accessible to all ranks.

\subsection{Microbenchmarks}\label{subsec:microbenchmarks}
We first test whether deriving collectives from one protocol (SNAC) retains the specialization needed for competitive performance (\cref{fig:eval-collectives}).
These four collectives use ordinary application buffers, so their results \emph{include} the work of internal staging.
We evaluate variable-length collectives in \cref{subsec:varlen-collectives}.

\para{Does SNAC deliver low latency?}
Across all collectives in \cref{fig:eval-collective-latency}, we improve latency over NCCL by 1.65--4.36$\times$ in geometric mean.
In AllReduce, \NAME achieves the largest latency improvement on each platform, reaching 4.36$\times$ on B200 while our largest individual speedup is 5.14$\times$ over NCCLX for 2\,KiB AllReduce on B200.

We attribute our low latency results to (1) our latency specializations (\cref{subsec:snac})
and (2) epoch-based double-buffering which eliminates inter-rank synchronization (\cref{subsec:implementation-epochs}).

MSCCL++, which also implements similar low-latency techniques, provides a closer comparison for AllGather and AllReduce.
Compared to MSCCL++, we improve AllGather latency by 1.23--1.56$\times$ in geometric mean across all platforms.
For AllReduce, our speedups range from 0.98$\times$ to 1.06$\times$.
These results show that collectives derived through SNAC match or exceed expert-developed implementations in MSCCL++ or NCCL.
\begin{figure}[!h]
  \centering
  \includegraphics[width=0.98\columnwidth]{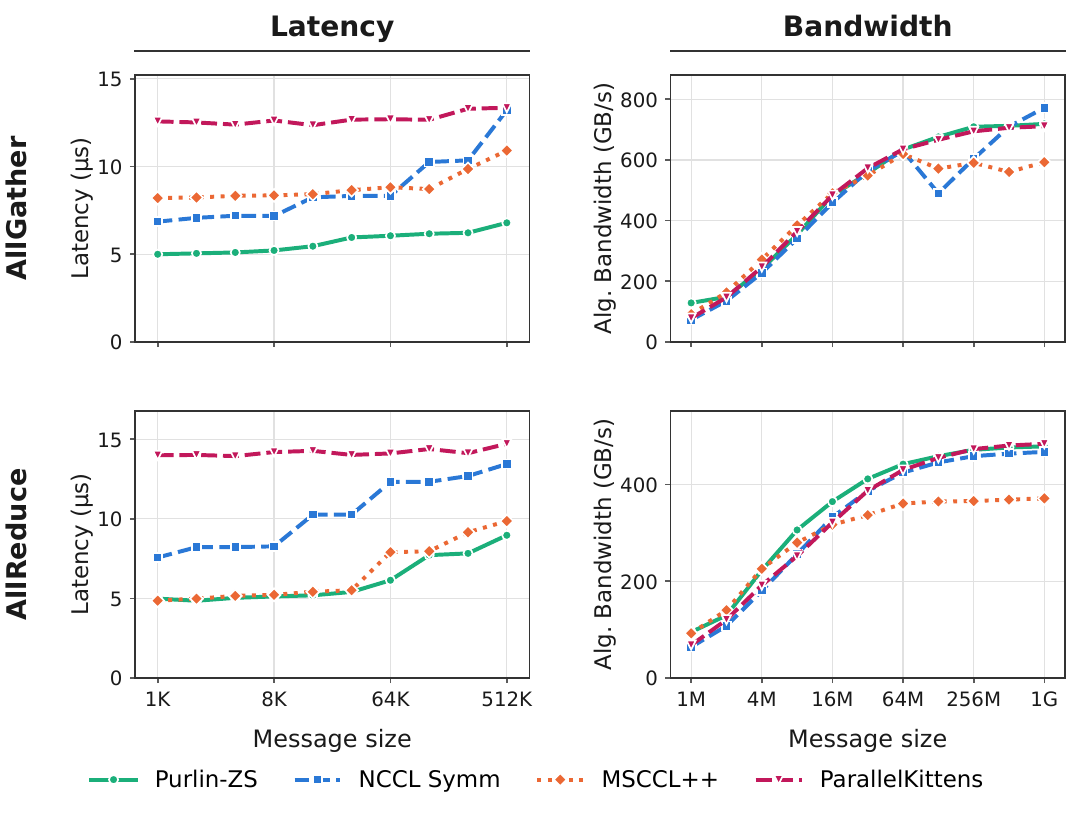}
  \par\vspace{3pt}
  \caption{Results on 8xB200 GPUs with peer-accessible buffers.}
  \label{fig:eval-symmetric-b200}
  \vspace{-12pt}
\end{figure}

\para{Does SNAC achieve high bandwidth?}
Across the four collectives in \cref{fig:eval-collective-bandwidth}, \NAME improves bandwidth, in geometric mean, over NCCL by up to 1.42$\times$ on A100, 1.25$\times$ on H200, and 1.53$\times$ on B200.
Compared to MSCCL++, which is the closer contender, \NAME improves bandwidth on AllReduce and AllGather by up to 1.24$\times$ on A100, 1.04$\times$ on H200 and 1.13$\times$ on B200.
As we show in \cref{subsec:ablation}, our speedups arise from (1) chunking within SNAC, which allows for overlapping staging with consuming, (2) the high-throughput pipeline of the \texttt{Atom},
and (3) tuned codesign policies.

We also investigate the performance of SCORE (\cref{subsubsec:score}): \NAME uses SCORE for \texttt{reduceScatter} on all three hardware generations and for \texttt{allReduce} on A100.
For \texttt{allReduce} on H200 and B200, \NAME uses NVLS (see \cref{app:multimem}) to stay competitive with baselines that also use NVLS.
On A100, we find that SCORE improves bandwidth over NCCL by 1.14--1.26$\times$ while on H200 and B200, we observe speedups of 1.05$\times$ and 1.35$\times$, respectively.

\para{Performance for peer-accessible buffers.}
We next compare systems using peer-accessible input and output buffers.
We see in \cref{fig:eval-symmetric-b200} that \NAME-ZS achieves 718.7\,GB/s for AllGather and 478.6\,GB/s for AllReduce, exceeding MSCCL++'s 592.4 and 371.1\,GB/s.
ParallelKittens reaches 710.9 and 483.9\,GB/s, respectively, but uses 148 SMs while \NAME-ZS uses only 32 SMs for both collectives.
For small messages, we improve latency over ParallelKittens by up to 2.9$\times$ and NCCL Symm by 1.70$\times$ across both AllGather and AllReduce.
We report results on A100 and H200 in \cref{app:peer-accessible}.
\begin{figure}[!h]
  \centering
  \includegraphics[width=\columnwidth]{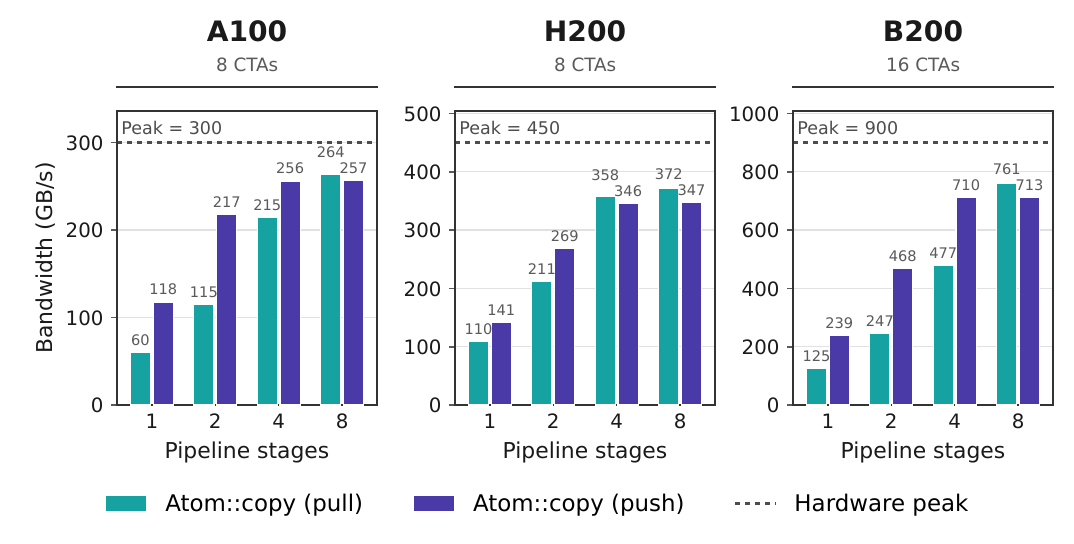}
  \captionsetup{skip=2pt}
  \caption{Copy bandwidth for push and pull over NVLink as we vary the pipeline depth of \texttt{Atom::copy}.}
  \vspace{-12pt}
  \label{fig:eval-pipeline-stages}
\end{figure}
\begin{figure}[!h]
  \centering
  \includegraphics[width=\columnwidth]{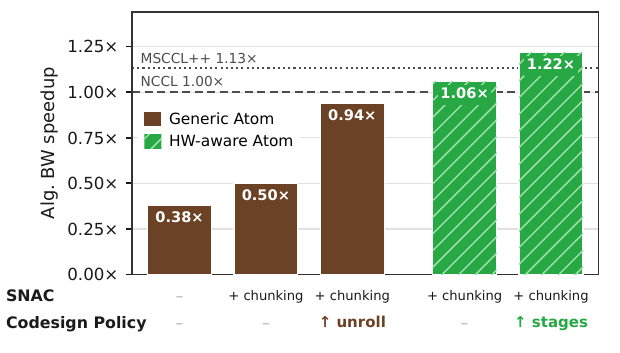}
  \captionsetup{skip=2pt}
  \caption{Ablation of \NAME's performance.}
  \label{fig:eval-ablation}
  \vspace{-12pt}
\end{figure}

\begin{figure*}[!t]
  \centering
  \includegraphics[width=\textwidth,trim=0 5bp 0 0,clip]{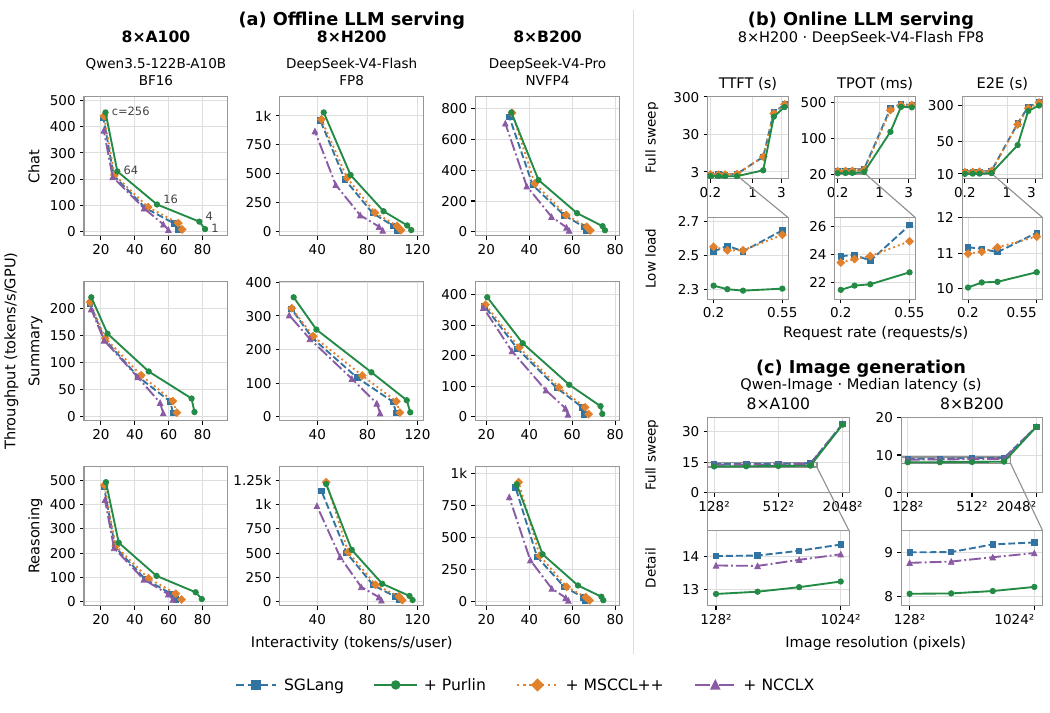}%
  \begingroup
  \phantomsubcaption\label{fig:eval-offline}
  \phantomsubcaption\label{fig:eval-online}
  \phantomsubcaption\label{fig:eval-diffusion}
  \endgroup
  \captionsetup{skip=3pt}
  \caption{End-to-end inference performance.
    (a) Offline LLM serving frontier. Rows use input/output lengths of 1,000/1,000 tokens (chat), 8,000/1,000 (summary), and 1,000/8,000 (reasoning).
    Interactivity is the inverse of median TPOT.
    (b) Median latencies of online serving on the Mooncake trace.
    (c) Median end-to-end latency for image generation. Both (a) and (b) use TP8/EP8/DP4 with data parallel attention.}
  \label{fig:eval-applications}
  \vspace{-12pt}
\end{figure*}

\subsection{Ablations}\label{subsec:ablation}
Here, we investigate how the unified pipeline improves bandwidth (\cref{subsubsec:pipeline-ablation}) and how components of \NAME together improve collective communication performance (\cref{subsubsec:purlin-ablation}).

\subsubsection{Impact of Pipeline Depth on Bandwidth}\label{subsubsec:pipeline-ablation}
We design the pipeline within the \ATOMTT to maximize bytes-in-flight per CTA toward increasing interconnect utilization (\cref{subsubsec:pipe-codesign}).
We therefore evaluate \emph{how} this mechanism improves bandwidth by varying pipeline depth while fixing the CTA (interchangeable with SM) allocation per GPU. Specifically, we set eight CTAs on A100 and H200 and sixteen on B200, each with 256 threads.
We further set $B_{\mathrm{stage}} = $ 8 KiB for A100 and $B_{\mathrm{stage}} = $ 16 KiB for H200 and B200 (see Listing~\ref{lst:atom_inst} for a breakdown of $B_{\mathrm{stage}}$).
Following \cref{eq:pipeline-window}, we sweep $Q_{\mathrm{cap}}$ by varying $S$ from 1 to 8.
For every hardware generation, $S=8$ is where $Q_{\mathrm{cap}} = $ Rounded BDP from \cref{tab:background-bdp}.

In \cref{fig:eval-pipeline-stages}, we observe that increasing depth from one to 8 stages raises pull and push bandwidth across generations on the \emph{same CTA allocation}.
We see up to a 6.1$\times$ bandwidth increase (125 to 761 GB/s) for pull on B200, while for push we see a 2.98$\times$ increase, with bandwidth improving from 239 to 713 GB/s.
These results confirm that maximizing bytes-in-flight indeed improves achieved interconnect bandwidth.

\subsubsection{How Components of Purlin Improve Performance}\label{subsubsec:purlin-ablation}
For this ablation, we focus on \texttt{allReduce} (ordinary buffers at 32 MiB), where chunking is important for performance (Algorithm~\ref{alg:allreduce}), and we run on A100s to control for hardware-accelerated NVLS, since we want to emphasize the efficacy of Purlin’s software techniques.

Our starting configuration is SNAC without chunking paired with a default codesign policy and \texttt{GenericAtom} (Listing~\ref{lst:atom_inst}) which uses vectorized data-movement instructions characteristic of pre-Ampere GPUs (\cref{subsec:evolving-hardware}). In \cref{fig:eval-ablation}, we see in the second bar that enabling chunking, which overlaps staging with consumption across \texttt{reduceScatter} and \texttt{allGather}, improves bandwidth by 1.31$\times$.
We then apply a codesign policy to increase the unroll depth of \texttt{GenericAtom}, increasing bytes-in-flight per SM.
This new policy yields 1.88$\times$ performance over the default policy and narrows the gap between NCCL to just 6\%.

Next, we swap \texttt{GenericAtom} (brown) for the hardware-aware \texttt{AmpereAtom} (green), which lowers to the \texttt{cp.async} unified pipeline in \cref{subsubsec:unified-pipeline}.
With \texttt{AmpereAtom}, \NAME outperforms NCCL by 6\% and improves bandwidth by 1.13$\times$ compared to the best performance of \texttt{GenericAtom}.
We finally tune \texttt{AmpereAtom} by increasing pipeline depth like in \cref{subsubsec:pipeline-ablation}.
As shown in the fifth bar of \cref{fig:eval-ablation}, applying this codesign policy allows \NAME to further outperform NCCL by 1.22$\times$ and execeed MSCCL++ by 1.08$\times$.

\para{Microbenchmark takeaway.}
In summary, our results validate that SNAC, from which we \emph{derive} all measured collectives, is \emph{compatible} across hardware generations and \emph{evolvable}, providing competitive performance for each generation.
Our results, particularly our ablation experiments, further confirm the utility of \emph{programmability}, which our \texttt{Atom} supports and which codesign policies provide.

\subsection{End-to-end Inference}\label{subsec:application-performance}
\textbf{Does \NAME impact end-to-end inference performance?}
We examine three settings: (1) offline LLM serving which varies the number of concurrent requests in flight,
(2) online LLM serving which varies the \emph{arrival rate} of requests, and
(3) image generation.
Together, they test \NAME across latency-sensitive (image and online LLM) and throughput-oriented (offline LLM) workloads.

\para{Offline LLM serving.}\label{subsec:llm-inference}
We configure SGLang with tensor and expert parallelism of eight, data parallelism of four, and data-parallel attention.
Thus from \cref{tab:background-parallelism}, collectives exercised in this experiment are AllGather(V), ReduceScatter(V), and AllReduce.
For each model in \cref{tab:eval-platforms}, we run three workloads, each at five concurrencies.
At concurrency $c$, we submit $8c$ requests with up to $c$ in flight.
We note that SGLang supplies ordinary application buffers to communication backends.

Across all baselines and workloads in \cref{fig:eval-offline}, SGLang+\NAME improves output throughput and interactivity by 1.13$\times$ in geometric mean and up to 1.37$\times$.
Per baseline, we observe geometric mean gains of 1.11$\times$ over stock SGLang, 1.08$\times$ over SGLang+MSCCL++, and 1.21$\times$ over SGLang+NCCLX.
These gains are particularly pronounced at low concurrency; for example, in the chat workload on A100, we improve interactivity over stock SGLang by 1.24$\times$ at concurrency one compared to 1.06$\times$ at concurrency 256.
This is consistent with the breakdown in \cref{fig:motivation} where communication accounts for a significant portion of end-to-end latency at low concurrency.
Improving collective communication performance in that regime can thus reduce end-to-end latency as we see for \NAME.
On the other hand, as concurrency increases, requests become more compute-bound, leaving little headroom, hence our gains narrow for high concurrency across all nine settings.

\para{Online LLM serving.}
Here, we investigate the setting where requests arrive at a varying rate.
We replay 1,000 requests from the Mooncake conversation trace~\cite{305212, mooncake_trace} on eight H200 GPUs with DeepSeek-V4-Flash FP8.
We vary request rates from 0.2 to 3.4 requests/s and measure median TTFT, TPOT, and end-to-end latency.

As shown in \cref{fig:eval-online}, we improve interactivity by 1.26$\times$ in geometric mean and up to 2.85$\times$.
At 0.2--0.55 requests/s, we improve median end-to-end latency over stock SGLang by 1.08--1.11$\times$.
At 1.5 requests/s, we reduce end-to-end latency from 121.3 to 41.5 seconds, a 2.92$\times$ speedup over stock SGLang and 2.78$\times$ over SGLang+MSCCL++.

We note that these larger gains occur under overload; specifically, we observe that achieved throughput plateaus at approximately 1.2--1.4 requests/s, thus queueing delay becomes more prominent for these and higher request rates.
We hypothesize that these larger gains stem from reduced queueing delay for backlogged requests.
That is, faster collectives reduce the latency of \emph{active requests}, which in turn reduces turnaround time for backlogged requests and therefore reduce the queueing delay portion of their total latency.

\para{Image generation.}\label{subsec:image-generation}
We run Qwen-Image~\cite{qwen_image} using Ulysses sequence parallelism~\cite{ulysses} of degree 8 which invokes \texttt{allToAll} across all 8 GPUs.
For each resolution in \cref{fig:eval-diffusion}, we issue eight requests at concurrency one and report median latency.
For $128^2$--$1024^2$, \NAME improves latency over stock SGLang by approximately 1.09$\times$ on A100 and 1.12--1.13$\times$ on B200.
We also outperform SGLang+NCCLX by 1.06--1.07$\times$ on A100 and approximately 1.09$\times$ on B200.
At $2048^2$, which is significantly compute-bound, all three systems perform similarly, as expected.

\subsection{Output Quality}\label{subsec:qualitative-evaluation}
\input{figures/e4_e5_quality}
We evaluate model quality to validate that SGLang + \NAME does not degrade model outputs.
On eight H200 GPUs, we observe in \cref{tab:quality} similar AIME26 accuracy with SGLang using DeepSeek-V4-Flash FP8.
We also observe identical scores for pass@16 and majority@16.
For Qwen-Image, we obtain pixel-identical outputs for all 100 prompts from Qwen-Image-Bench~\cite{qwen_image_bench} at $1024^2$ resolution.

%% file: figures/e4_e5_quality.tex
\begin{table}[!h]
  \centering
  \small
  \begin{tabular}{@{}lrr@{}}
    \toprule
    Metric & SGLang & + \NAME \\
    \midrule
    \multicolumn{3}{@{}l}{\emph{AIME26: 30 problems, 16 samples each}} \\
    pass@1 (\%, mean $\pm$ SEM) & $96.04\pm0.34$ & $95.62\pm0.40$ \\
    pass@16 (\%) & 96.67 & 96.67 \\
    majority@16 (\%) & 96.67 & 96.67 \\
    \midrule
    \multicolumn{3}{@{}l}{\emph{Qwen-Image: 100 prompts}} \\
    Pixel-identical image pairs & \multicolumn{2}{c}{100 / 100} \\
    Mean squared pixel error & \multicolumn{2}{c}{0} \\
    \bottomrule
  \end{tabular}
  \caption{Output quality. SEM is standard error of the mean.}
  \label{tab:quality}
\end{table}

%% file: sections/6_discussion.tex
\section{Limitations and Future Work}\label{sec:limitations}
\para{Does \NAME scale beyond eight GPUs?}
We evaluate only eight-GPU NVLink domains, the largest scale-up systems accessible to us.
Larger scale-up domains such as NVL72 may demand different implementation techniques to achieve peak performance, while multi-node RDMA settings would need a scale-out \texttt{Atom}.
We leave both to future work.

\para{Which hardware mechanisms are compatible with the \texttt{Atom}?}
\emph{Device-initiated} mechanisms such as \texttt{cp.async} and TMA interoperate with the \texttt{Atom}, whereas
host-initiated mechanisms such as the copy engine are incompatible.
We believe future work may involve developing device-side interfaces for these host-side mechanisms.

\para{Implementations for future work?}
In \Cref{app:rooted-collectives}, we show how to express rooted collectives (Broadcast, Reduce, Scatter, and Gather) in SNAC. However, our software prototype implements only non-rooted collectives, which we evaluated in \cref{subsec:microbenchmarks} and \cref{subsec:varlen-collectives}.
We leave implementing and tuning these rooted collectives for future work.
Another relevant direction is integrating our \texttt{Atom} into fused distributed kernels such as FlashMoE~\cite{NEURIPS2025_918d938b}, MegaMoE~\cite{deepseekai2026} or MPK~\cite{318585}.

\section{Related Work}\label{sec:related-work}
\textbf{Collective synthesis and compilation.}
SCCL~\cite{cai2021sccl} synthesizes collective algorithms for a given topology, and TACCL~\cite{shah2023taccl} uses communication sketches to guide synthesis.
OptCCL~\cite{shapley2026optccl} scales synthesis to hundreds of GPUs and supports joint optimization of concurrent collectives.
In contrast, we focus on developing a shared orchestration protocol that allows for \emph{deriving} hardware-aware collectives from a compile-time description.

\para{Compatibility in GPU communication.}
In pursuit of compatibility and extensibility, UCCL-Tran~\cite{uccl_transport} separates the control path from the datapath of GPU-initiated RDMA communication, while UCCL-EP~\cite{uccl_ep} applies a similar separation to expert-parallel communication over RDMA.
We apply a similar decoupling principle to communication within a \emph{scale-up domain}.
Our separation of orchestration from the datapath of collectives lets us evolve the datapath without rewriting orchestration or sacrificing compatibility.

\para{Composable GPU abstractions.}
CUTLASS and CuTe~\cite{nvidia_cutlass3} decompose high-performance compute operations, notably GEMMs, into reusable hardware-aware components manipulated via expressive layouts~\cite{cecka2026cutelayoutrepresentationalgebra}.
We draw inspiration from these works; specifically, our layouts enable us to parametrize SNAC to yield high-performance collectives.

%% file: sections/7_conclusion.tex
\section{Conclusion}\label{sec:conclusion}

We presented the design and implementation of \NAME, a scale-up communication framework that decouples orchestration from the datapath of collective communication.
In this decoupled stack, we express collectives through layouts, derive orchestration through the SNAC protocol, and realize hardware-aware data movement via the \texttt{Atom}.
Our evaluation shows that this separation yields competitive or better performance for collectives and distributed inference across multiple GPU generations.
\NAME demonstrates that it is indeed possible for us to have a communication framework that can \emph{keep pace} with evolving applications and hardware while preserving performance and programmability.

\section{Acknowledgements}\label{sec:acknowledgements}
OJA gratefully acknowledges support from the National Science Foundation Graduate Research Fellowship Program (NSF GRFP) under Grant No. DGE-2146755.
Any opinions, findings, and conclusions or recommendations expressed in this paper are those of the authors and do not necessarily reflect the views of the National Science Foundation.

%% file: sections/appendix.tex
\appendix
\twocolumn[{
  \begin{minipage}{\textwidth}
    \input{figures/rooted_collectives}
  \end{minipage}
  \vspace{\baselineskip}
}]
\section{Rooted Collectives in SNAC}\label{app:rooted-collectives}
We can describe rooted collectives, namely \texttt{broadcast}, \texttt{reduce}, \texttt{scatter}, and \texttt{gather} with our existing layouts by further describing the role of the root within the SNAC layout transformation.
Precisely, we propose adding a compile-time \texttt{Root} parameter to SNAC that captures whether a root is a producer, as in \texttt{scatter}, or a consumer, as in \texttt{gather}.
Listing~\ref{lst:rooted-collectives} depicts \texttt{Root} and how we use this parameter to express \texttt{scatter} and \texttt{gather} as SNAC transformations.

\para{Composing rooted transformations.}
Bandwidth-optimal implementations for \texttt{broadcast} and \texttt{reduce} are known as compositions; specifically, \texttt{broadcast} is \texttt{scatter} then \texttt{allGather} while \texttt{reduce} is \texttt{reduceScatter} followed by \texttt{gather}.
We leverage our \texttt{compose} primitive to express these compositions in SNAC as shown in Listing~\ref{lst:rooted-collectives}.
\section{SCORE vs. NVLink SHARP}\label{app:multimem}
\begin{figure}[!h]
  \centering
  \includegraphics[width=0.8\columnwidth]{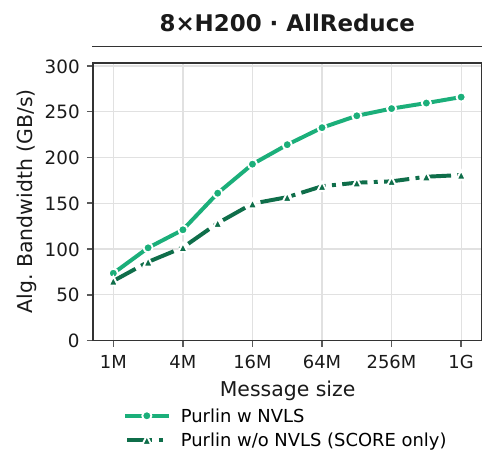}
  \caption{\NAME's AllReduce performance comparing SCORE (\cref{subsubsec:score}) and NVLink SHARP~\cite{9895480, hopper_whitepaper}. Higher is better.}
  \label{fig:score_v_nvls}
\end{figure}
\begin{table}[!h]
  \vspace{-10pt}
  \centering
  \footnotesize
  \setlength{\tabcolsep}{5pt}
  \begin{tabular}{@{}lccc@{}}
    \toprule
    Technique & \makecell{Achieved\\Alg. Bandwidth} & \makecell{Peak\\Alg. Bandwidth} & \% of peak\\
    \midrule
    SCORE & 180.7\,GB/s & $\frac{n}{2(n - 1)} \times 450 = 257$ GB/s & $\frac{180.7}{257} = $ \textbf{70\%} \\
    NVLS & 265.9\,GB/s & $\frac{n}{n + 1} \times 450 = 400$ GB/s & $\frac{265.9}{400} = $ \textbf{66\%} \\
    \bottomrule
  \end{tabular}
  \caption{Percentage of corresponding peak Alg. BW achieved by SCORE and NVLS. Achieved Alg. BW is the bandwidth measured at 1\,G. Note that $n=8$ and 450 GB/s is the nominal unidirectional bandwidth for H200.}
  \label{tab:score_v_nvls_tab}
\end{table}
Here, we compare the performance of SCORE and NVLS.
In \cref{fig:score_v_nvls}, we show the algorithm bandwidth of AllReduce in \NAME with and without NVLS.
We observe there that \NAME with NVLS improves algorithm bandwidth by up to 1.47$\times$ compared to SCORE.
This performance gap is due to algorithmic differences.
Specifically, \NAME without NVLS executes AllReduce as a ReduceScatter followed by an AllGather, so for message size $S$ and $n$ ranks, each rank sends $\frac{2(n-1)}{n}S$ bytes.
With NVLS, Purlin also performs a ReduceScatter and AllGather but both are performed via multicast mechanisms by NVSwitch; the resulting bandwidth cost is thus lowered to $\frac{n+1}{n}S$~\cite{9895480}.
At $n=8$, these costs are $1.75S$ for SCORE and $1.125S$ for NVLS. Thus, SCORE moves 1.56$\times$ more bytes than NVLS, implying a higher peak algorithm bandwidth for NVLS than for SCORE, as we show in \cref{tab:score_v_nvls_tab}.

As a result, the choice between these two techniques is a tradeoff: NVLS delivers higher bandwidth but does not guarantee the deterministic reduction that SCORE provides.
We therefore expose both techniques in \NAME, deferring the final choice to the application.

\section{Additional Evaluation Details}\label{app:evaluation-details}
\subsection{Microbenchmark methodology}\label{subsec:microbenchmark-methodology}
For all microbenchmarks, we capture 128 invocations in a CUDA graph, replay that graph to warm up, and time eight replays with CUDA events.
We divide each rank's elapsed time by 128 $\times$ 8 = 1,024 and report the maximum across ranks.
We use BF16 as the datatype for all collectives involving reductions.
Reported message size is the input size for ReduceScatter and the output size for AllGather, AllToAll, and AllReduce.
We compute algorithm bandwidth as message size divided by latency.
\subsection{End-to-end inference configuration}\label{subsec:software-and-application-configuration}
All end-to-end integrations use SGLang revision \texttt{0f18d38}, with reported version 0.5.19.dev806+g0f18d389b.
We use PyTorch 2.13.0 and FlashInfer 0.6.18.
For online serving, we use the first 1,000 Mooncake conversation requests.
For image generation, we use Ulysses sequence parallelism across eight GPUs, with ring parallelism disabled (degree one).
For our output quality experiment on AIME26, we enable thinking, sample at temperature 1.0 and top-$p$ 0.95 and permit up to 131,072 output tokens.

\subsection{Additional Peer-Accessible Buffer Results}\label{app:peer-accessible}
We report results for peer-accessible buffers on A100 and H200 below.
\begin{figure}[!h]
  \centering
  \includegraphics[width=0.98\columnwidth]{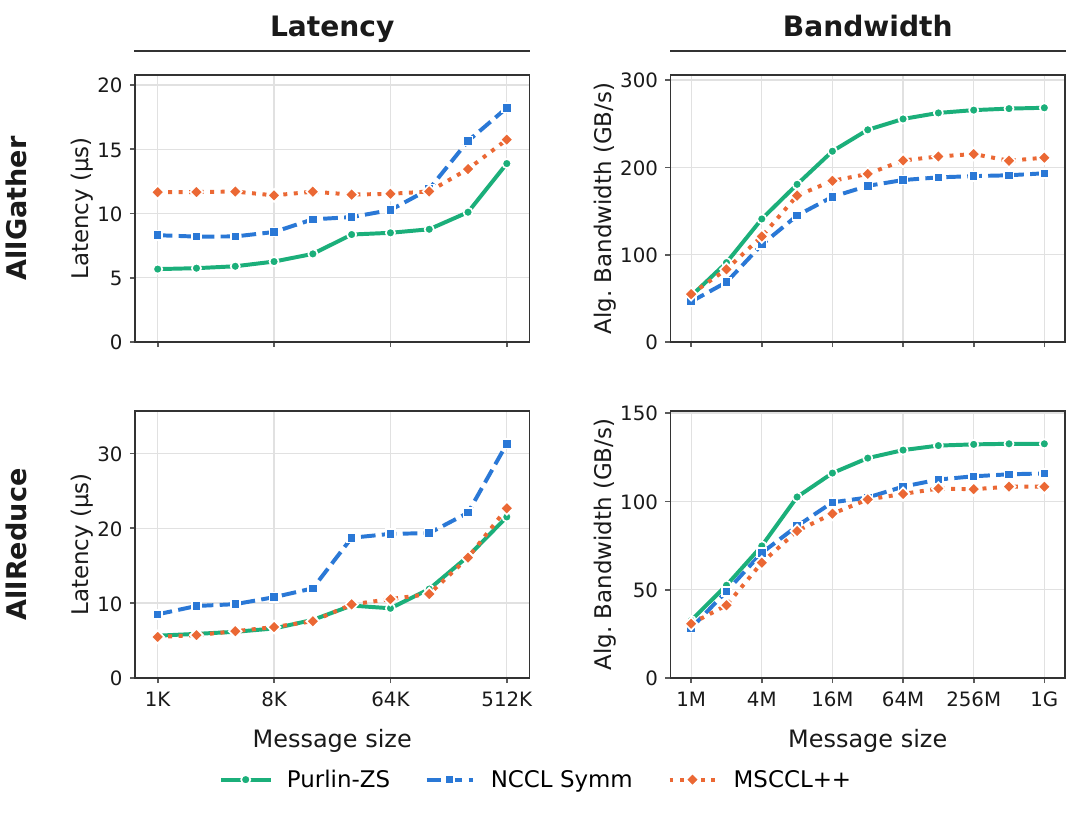}
  \caption{Results on 8$\times$A100 GPUs.}
  \label{fig:eval-symmetric-a100}
\end{figure}

\begin{figure}[!h]
  \centering
  \includegraphics[width=0.98\columnwidth]{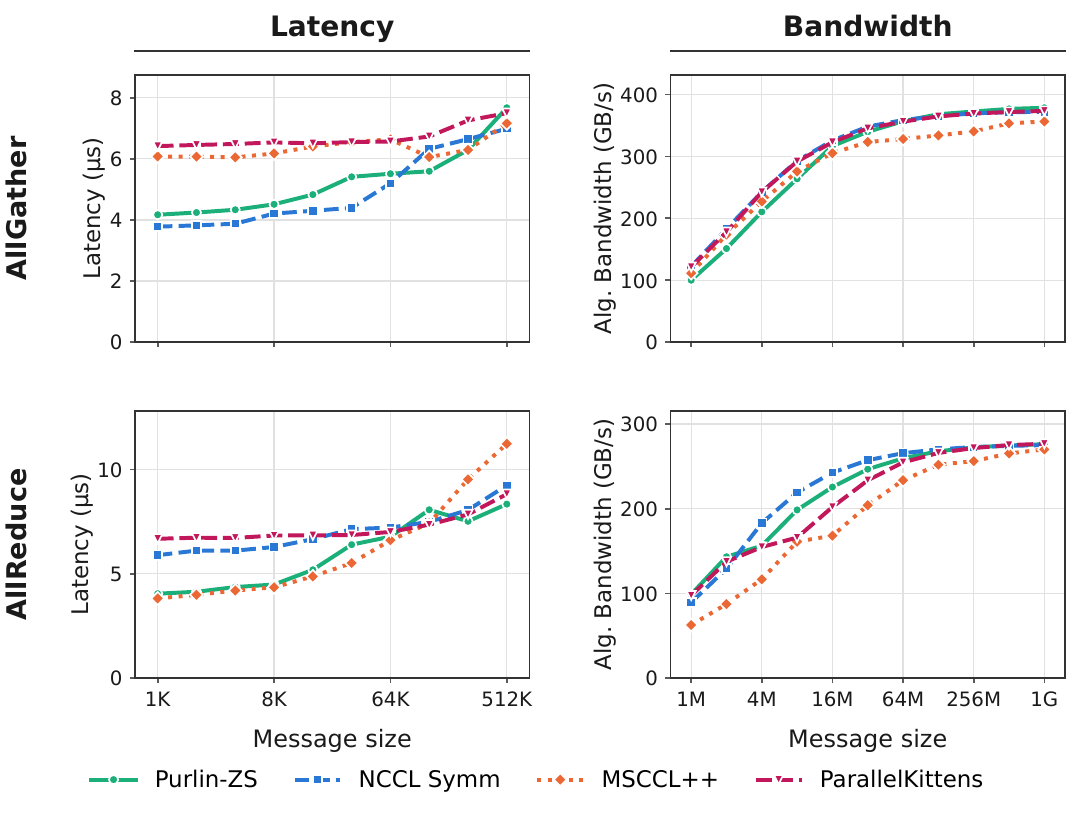}
  \caption{Results on 8$\times$H200 GPUs.}
  \label{fig:eval-symmetric-h200}
\end{figure}

\twocolumn[{
  \begin{minipage}{\textwidth}
  \captionsetup{type=figure}
  \centering
  \begin{subfigure}[t]{0.495\textwidth}
    \centering
    \includegraphics[width=\linewidth]{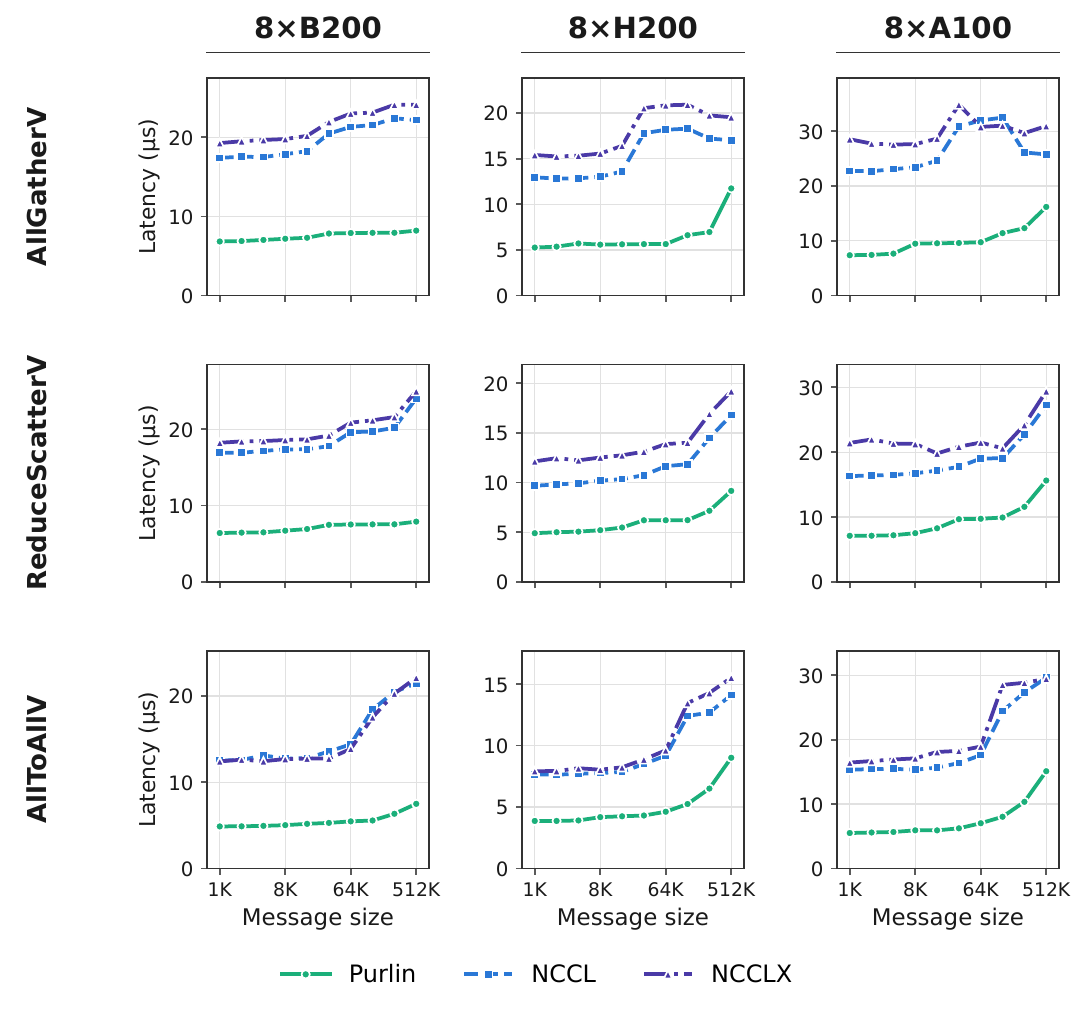}
    \caption{Latency. Lower is better.}
    \label{fig:eval-collective-v-latency}
  \end{subfigure}\hfill
  \begin{subfigure}[t]{0.495\textwidth}
    \centering
    \includegraphics[width=\linewidth]{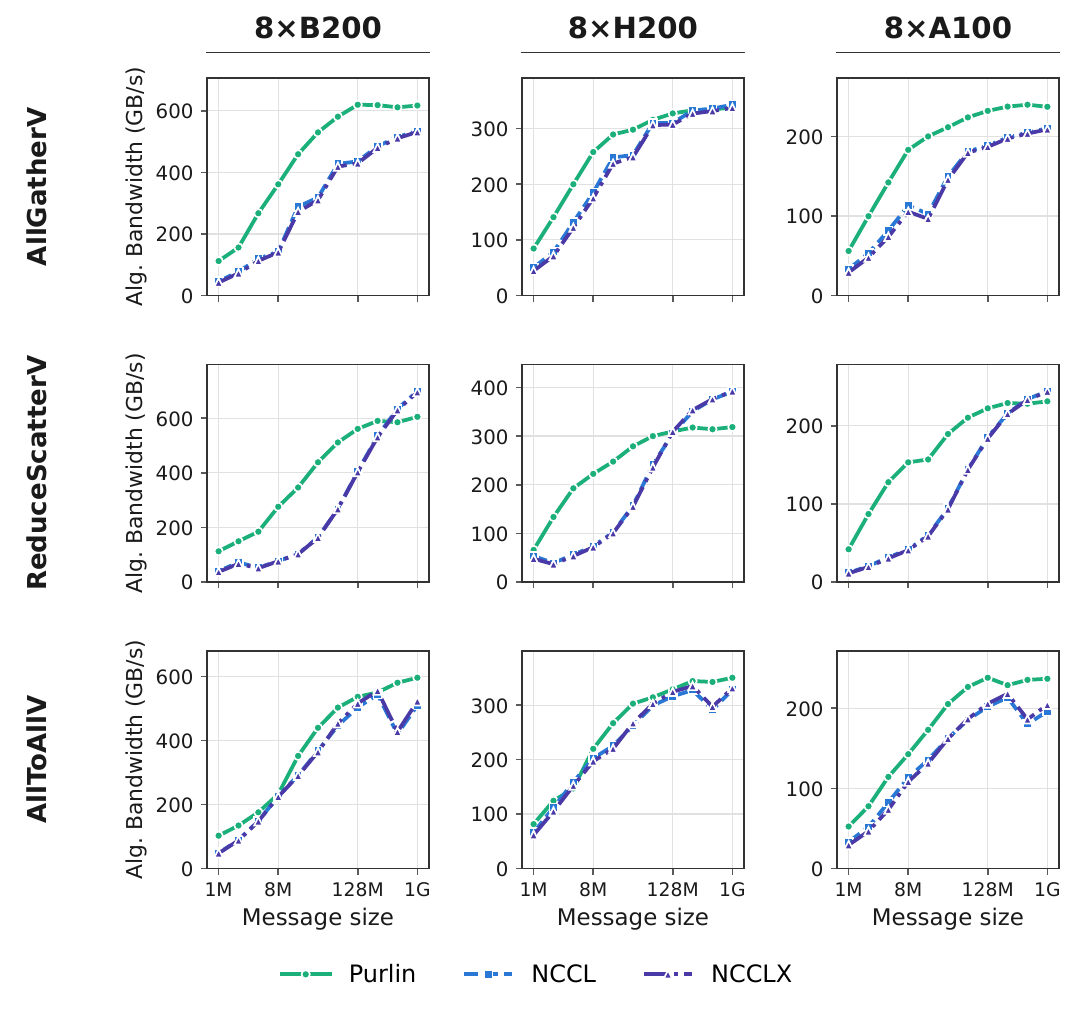}
    \caption{Algorithm bandwidth. Higher is better.}
    \label{fig:eval-collective-v-bandwidth}
  \end{subfigure}
  \caption{Variable-length collectives with ordinary buffers. Message size is $T$ in \cref{eq:zipf-split} and partitioned via Zipf with exponent $s=0.125$.}
  \label{fig:eval-collectives-v}
  \end{minipage}
  \vspace{\baselineskip}
}]

\subsection{Variable-Length Collectives}\label{subsec:varlen-collectives}
We evaluate AllGatherV, ReduceScatterV, and AllToAllV against NCCL and NCCLX on all three platforms, using the ordinary buffers and methodology from \cref{sec:evaluation}.
We note that partition sizes are passed to these collectives at runtime.
NCCL does not provide native implementations for these collectives, so we implement them using existing primitives, following current implementations in SGLang and vLLM, as follows:
\begin{itemize}
  \item \textbf{AllGatherV:} one \texttt{ncclBroadcast} per rank, which copies that rank's partition into every rank's output.
  \item \textbf{ReduceScatterV:} one \texttt{ncclReduce} per rank, which reduces that rank's partition from every rank's input into that rank's output.
  \item \textbf{AllToAllV:} one \texttt{ncclSend} and one \texttt{ncclRecv} per peer, plus a local copy of the rank's own partition.
\end{itemize}
Each emulation issues its calls in a single NCCL group (\texttt{ncclGroupStart}/\texttt{ncclGroupEnd}).
For NCCLX, we call the variable-length APIs of TorchComms: \texttt{all\_gather\_v}, \texttt{reduce\_scatter\_v}, and \texttt{all\_to\_all\_v\_single}.

\para{Variable-length splits.}
We assign $n_r$ bytes to rank $r$ from a total of $T$ bytes across $W=8$ ranks using a Zipf split with exponent $s=0.125$:
\begin{equation}
  \begin{aligned}
    x_r &= \frac{T}{u}\frac{(r+1)^{-s}}{\sum_{k=0}^{W-1}(k+1)^{-s}}, \\
    n_r &= u\bigl(\lfloor x_r\rfloor+\delta_r\bigr).
  \end{aligned}
  \label{eq:zipf-split}
\end{equation}
For our power-of-two totals, we use $u=\max(128,T/(8W))$ bytes and assign the remaining units through $\delta_r\in\{0,1\}$ in descending order of the fractional parts of $x_r$, breaking ties toward lower ranks.
This preserves the total with a largest partition of $1.125\times$ the mean of all partitions.

AllGatherV uses $n_r$ as rank $r$'s contribution, whereas ReduceScatterV reduces $T$ bytes per rank into an $n_r$-byte output at rank $r$.
For AllToAllV, we rotate the partition vector right by $(r+1)\bmod W$ positions at rank $r$, so every rank sends and receives $T$ bytes.
We plot $T$ for all three collectives in \cref{fig:eval-collectives-v} and use identical partitions for baselines as well.

\para{Performance.}
In \cref{fig:eval-collective-v-latency}, we see that \NAME improves latency compared to NCCL by up to 2.73$\times$ and NCCLX by up to 3.03$\times$ in geometric mean across all platforms and collectives.
In \cref{fig:eval-collective-v-bandwidth}, we see that \NAME improves geometric mean bandwidth over NCCL by up to 1.65$\times$ for AllGatherV, 1.29$\times$ for AllToAllV, and 2.00$\times$ for ReduceScatterV.
ReduceScatterV provides the largest individual gain: at 2\,MiB on A100, we achieve 87.0\,GB/s against NCCL's 20.4\,GB/s and NCCLX's 19.3\,GB/s, improvements of 4.26$\times$ and 4.50$\times$.
At 1\,GiB, however, NCCL leads ReduceScatterV on all three platforms.
The largest gap occurs on H200, where we achieve 318.5\,GB/s against NCCL's 391.4\,GB/s.

%% file: figures/rooted_collectives.tex
\begin{listing}[H]
\begin{minted}{cpp}
using enum ConsumeOp;
using enum DataLayout;
enum class Root { none, producer, consumer };

using allGather      = SNAC<Atom, Policy, copy,   packed,     scattered, Root::none>;
using reduceScatter  = SNAC<Atom, Policy, reduce, scattered,  packed,    Root::none>;
// rooted collectives
using scatter        = SNAC<Atom, Policy, copy,   scattered,  packed,    Root::producer>;
using gather         = SNAC<Atom, Policy, copy,   packed,     scattered, Root::consumer>;
using broadcast      = compose<scatter, allGather>;
using reduce         = compose<reduceScatter, gather>;
\end{minted}
\caption{Proposed declarations of rooted collectives.}
\label{lst:rooted-collectives}
\end{listing}